\documentclass[12pt, a4paper]{article}
\usepackage{cite}
\usepackage{amsmath,amssymb}
\input{colordvi.tex}
\usepackage{pifont}
\usepackage{subfigure}
\usepackage{comment}
\usepackage{bm}
\usepackage{url}
\usepackage{braket}

\usepackage[linktocpage]{hyperref}
\usepackage{ytableau}
\everymath{\displaystyle}
\usepackage{cancel}
\usepackage{mathtools}
\usepackage{tabularx}
\usepackage{lipsum}

\long\def\symbolfootnote[#1]#2{\begingroup%
  \def\thefootnote{\fnsymbol{footnote}}\footnote[#1]{#2}\endgroup}

\usepackage{ifpdf}
\ifpdf
  \usepackage{graphicx, hyperref, xcolor}     %   usepackage without driver option
\else     % For (p)LaTeX + dvipdfmx
  \usepackage[dvipdfmx]{graphicx, hyperref, xcolor}     %   usepackage with driver option
 \fi

\newcommand{\be}{\begin{equation}}
\newcommand{\ee}{\end{equation}}
\newcommand{\ba}{\begin{eqnarray}}
\newcommand{\ea}{\end{eqnarray}}
\newcommand{\vev}[1]{\langle #1 \rangle}

\definecolor{blue}{rgb}{0, 0.4, 0.7}
\definecolor{Blue}{rgb}{0, 0, 1}
\hypersetup{% hyperref option list
setpagesize=false,
bookmarksnumbered=true,%
bookmarksopen=true,%
colorlinks=true,%
linkcolor=blue,
urlcolor=blue,
citecolor=blue,
}

\DeclareUnicodeCharacter{2212}{-}
\begin{document}
\fontsize{12pt}{18pt}\selectfont

%%%%%%%%%%%%%%%%%%%%%%%%%%%%%%%%%%%%%%%%%%%%%%%%%%
\begin{titlepage}

\begin{center}

\hfill CERN-TH-2023-023\\

\vskip .75in

{\Large \bf
$Z_{2}$-Odd Polonyi Field in Twin Higgs Model}

\vskip .75in

{\large
Gongjun Choi\symbolfootnote[1]{gongjun.choi@cern.ch}$^{1}$ and Keisuke Harigaya\symbolfootnote[2]{kharigaya@uchicago.edu}$^{1,2,3,4,5}$
}
\vskip 0.25in

$^1${\em Theoretical Physics Department,
      CERN, 1211 Geneva 23, Switzerland}\\[.3em]
$^2${\em Department of Physics, University of Chicago, Chicago, IL 60637, USA}\\
      [.3em]
$^3${\em Enrico Fermi Institute, University of Chicago, Chicago, IL 60637, USA}\\
      [.3em]
$^4${\em Kavli Institute for Cosmological Physics, University of Chicago, Chicago, IL 60637, USA}\\
      [.3em]
$^5${\em Kavli Institute for the Physics and Mathematics of the Universe (WPI),\\
The University of Tokyo Institutes for Advanced Study,\\
The University of Tokyo, Kashiwa, Chiba 277-8583, Japan}\\
      [.3em]
\vskip 0.20in

\end{center}
\vskip .5in

\begin{abstract}
We consider a supersymmetric mirror Twin Higgs model in a gravity-mediated supersymmetry-breaking scenario. We point out that the Polonyi field can be odd under the $Z_{2}$ symmetry exchanging the Standard Model with the mirror sector while gaugino masses are generated at tree-level. We discuss the dynamics of the Polonyi field during and after inflation and show that the Polonyi problem is absent. The Polonyi field couples to the two sectors with opposite signs, which may serve as origin of the $Z_{2}$-breaking of the Higgs potential in Twin Higgs models. We also estimate the $Z_2$-breaking in soft masses of supersymmetric particles.
\end{abstract}

\end{titlepage}

%\tableofcontents

\renewcommand{\thepage}{\arabic{page}}
\setcounter{page}{1}
\renewcommand{\thefootnote}{\arabic{footnote}}
\setcounter{footnote}{0}
\renewcommand{\theequation}{\thesection.\arabic{equation}}

%%%%%%%%%%%%%%%%%%%%%%%%%%%%%%%%%%%%%%%%%%%%%%%%%%

\newpage

\tableofcontents

\newpage

%%%%%%%%%%%%%%%%%%%%%%%%%%%%%%%%%%%%%%%%%%%%%%%%%%
\section{Introduction}
\label{sec:Intro}
\setcounter{equation}{0}
%%%%%%%%%%%%%%%%%%%%%%%%%%%%%%%%%%%%%%%%%%%%%%%%%%
In mirror Twin Higgs (TH) theories, the exact copy of gauge groups and particle contents of the SM sector are introduced, and two sectors so obtained enjoy $Z_{2}$ exchanging symmetry. Due to $Z_{2}$, there arises an accidental global $SU(4)$ symmetry of the mass terms of the Higgses.
Further assuming that the Higgs quartic terms are also approximately $SU(4)$ symmetric, the SM-like Higgs boson can be understood as the pseudo Nambu-Goldstone boson (pNGB) of the spontaneously broken global $SU(4)$ by the twin Higgs~\cite{Chacko:2005pe,Barbieri:2005ri}. As such, the quadratic sensitivity of the SM-like Higgs mass squared to the UV cut-off scale is relaxed, and so is the little hierarchy problem of the electroweak scale. The two main challenges of the TH theories concern 1) the $SU(4)$ invariant quartic coupling and
2) the big hierarchy between the twin electroweak (EW) scale  and the fundamental scale.

These two challenges may be resolved by extending TH theories by supersymmetry (SUSY).
The twin EW scale can be set by the soft mass scale, and the $SU(4)$ invariant quartic can be obtained by an $F$-term potential from a new singlet field~\cite{Falkowski:2006qq,Chang:2006ra,Craig:2013fga,Katz:2016wtw} or a $D$-term potential from a new gauge interaction~\cite{Badziak:2017syq, Badziak:2017kjk,Badziak:2017wxn}. The SUSY TH models are indeed able to explain the EW scale with a mild tuning. For example, in the $D$-term model with an $U(1)$ gauge symmetry~\cite{Badziak:2017syq}, the required tuning can be as low as $\sim20\%$. Another noticeable thing in SUSY TH models is that the tree-level SM-like Higgs mass is larger than the MSSM one by about a factor $\sqrt{2}$, and the observed SM-like Higgs boson mass can be obtained without radiative corrections from a large stop mass, which also reduces fine-tuning.
As a further bonus, the lightest SUSY particle, which may be a super-partner of a mirror particle~\cite{Badziak:2019zys,Badziak:2022eag}, is a dark matter candidate~\cite{Witten:1981nf,Pagels:1981ke,Goldberg:1983nd}.

On the other hand, the simplest scheme of the mediation of SUSY breaking, namely, gravity mediation, suffers from the infamous Polonyi problem~\cite{Coughlan:1983ci}.
In gravity mediation, gaugino masses of the order of scalar masses require a singlet SUSY-breaking field called the Polonyi field. Although the Polonyi field $S$ may sit at a minimum of the potential during inflation thanks to a Hubble-induced potential, that minimum is in general displaced from the minimum after inflation by as large as $\mathcal{O}(M_{P})$,  where $M_{P}\simeq2.4\times10^{18}{\rm GeV}$ is the reduced Planck mass. The resultant oscillation of $S$ around the minimum is dangerous as it (or its decay products)  overcloses the universe afterward. The problem is fundamentally attributed to the lack of any charge of $S$ and hence the absence of the symmetry enhanced point of the $S$ field space. A solution to the Polonyi problem by a large coupling of the Polonyi field with the inflaton~\cite{Linde:1996cx,Takahashi:2011as,Nakayama:2012mf} and that by a coupling with a pseudo-flat direction~\cite{Harigaya:2013ns} have been considered in the literature.

Motivated by the advantages of SUSY TH models and the cosmological danger lurking in the gravity mediation, we consider the possibility where $S$ transforms as $S\rightarrow-S$ under the $Z_{2}$ symmetry exchanging the SM and mirror sectors. Such a transformation rule of $S$ clearly pins down the origin of the field space of $S$, providing a complete solution to the Polonyi problem. Moreover, in this set-up, any neutral (up to $Z_2$) operator $\mathcal{O}$ in the SM and $\mathcal{O}'$ in the mirror SM can couple to $S$ via $S(\mathcal{O}-\mathcal{O}')$. $\mathcal{O}$ includes gauge kinetic terms, so tree-level gaugino masses are obtained despite the $Z_2$-odd charge of $S$. As we will see, the spontaneous $Z_2$ breaking by the $F$ term of $S$, together with R symmetry breaking, can contribute to the phenomenologically required $Z_{2}$-breaking terms in the Higgs potential either at tree-level or by  radiative corrections.

The outline of the paper is as follows. In Sec.~\ref{sec:susyth}, we briefly review the Higgs sector in the SUSY TH models. Sec.~\ref{sec:model} is dedicated to discussion for the generation of gaugino masses, trilinear scalar couplings, scalar masses, and the $\mu$ and $b$ terms in the presence of the $Z_{2}$ odd field $S$. We show that the $Z_{2}$ breaking in the Higgs potential can be sufficiently suppressed. In Sec.~\ref{sec:mHZ2}, we compute the SM-like Higgs mass. Finally in Sec.~\ref{sec:Polonyi}, we present a concrete model for the SUSY breaking sector and study the early-universe dynamics of $S$.

We use the same symbol for a chiral superfield and its scalar component. We also use primes to denote the fields in the mirror sector. Finally, the R-charge of an operator $\mathcal{O}$ is denoted as $R[\mathcal{O}]$. 

%%%%%%%%%%%%%%%%%%%%%%%%%%%%%%%%%
\section{Review of Supersymmetric Twin Higgs}
\label{sec:susyth}
\setcounter{equation}{0}
%%%%%%%%%%%%%%%%%%%%%%%%%%%%%%%%%
In this section, we briefly review SUSY TH models.

As in generic TH models, the $Z_2$ symmetry exchanging the SM particles with their mirror partners is introduced. The $Z_2$ symmetry is extended to SUSY particles, so the mirror squarks, sleptons, gauginos, and higgsinos are also introduced.

The Higgs sector at the minimal contains $H_u$, $H_d$, $H_u'$, and $H_d'$, and they have quartic and quadratic terms. The quartic terms arise from the $F$ and $D$ terms. To be concrete, we only introduce $D$ term potentials from the SM and mirror electroweak symmetry,
\begin{align}
    V_{D,{\rm EW}}=\left(\frac{g_{2}^{2}+g_{Y}^{2}}{8}\right)\left[\left(|H_{u}|^{2}-|H_{d}|^{2}\right)^{2}+\left(|H'_{u}|^{2}-|H'_{d}|^{2}\right)^{2}\right] ,
\end{align}
and that from a new gauge symmetry under which both the SM and mirror particles are charged. The form of this potential depends on models~\cite{Badziak:2017syq, Badziak:2017kjk,Badziak:2017wxn}. To be concrete, we assume a gauge symmetry with an opposite charge of $H_d$ to $H_u$. The new gauge symmetry is broken at a scale above the SM and mirror EW symmetry breaking scale. If this occurs via a supersymmetric potential, the $D$ term potential of the new gauge interaction decouples. A non-zero $D$ term remains if the gauge symmetry breaking involves a SUSY-breaking potential, which is parameterized by $\epsilon$,
\begin{align}
    V_{D,{\rm new}} = \left(\frac{\lambda_X}{2}\right)^2 \left( |H_u|^2 - |H_d|^2 + |H_u'|^2 - |H_d'|^2 \right)^2,~~\lambda_{X}^{2}\equiv(1-\epsilon^{2})g_{X}^{2}/2,
 \label{eq:VDnew}   
\end{align}
where $g_X$ is the new gauge coupling constant.
The $D$ term potential from the EW symmetry violates the $SU(4)$ symmetry but preserves the $Z_2$ symmetry, while the one from the new gauge interaction preserves both $SU(4)$ and $Z_2$. 

Quadratic terms arise from a supersymmetric mass $\mu$ and soft supersymmetry breaking,
\begin{align}
    V_2 =& \left( |\mu^2| + m_{H_u}^2 \right) |H_u|^2  + \left( |\mu^2| + m_{H_d}^2 \right) |H_d|^2 +  \left( |\mu^{'2}| + m_{H_u'}^2 \right) |H_u'|^2 +  \left( |\mu^{'2}| + m_{H_d'}^2 \right) |H_d'|^2 \nonumber \\
    & + \left( b H_u H_d + {\rm h.c.} \right) + \left( b' H_u' H_d' + {\rm h.c.} \right).
\end{align}
As we will see in the next section, $Z_2$-breaking masses arise at tree-level or from quantum corrections.
Without loss of generality, we assume $Z_2$ breaking such that $v^2 = |\vev{H_u}|^2 + |\vev{H_d}|^2 < v^{'2} = |\vev{H_u'}|^2 + |\vev{H_d'}|^2  $.
In order for the potential to be bounded from below, $2|b'|<m_{H_{u}'}^{2}+m_{H_{d}'}^{2}+2|\mu'|^{2}$ is required.

We work in the decoupling limit of heavy Higgses so that the effective theory around the mirror and SM EW scale is given by that of the SM-like Higgs $H$ and its partner $H'$, which may be obtained by the following replacement;
\be
H_{u}=H\sin\beta\quad,\quad H_{d}=H\cos\beta\quad,\quad H_{u}'=H'\sin\beta'\quad,\quad H'_{d}=H'\cos\beta'\,,
\label{eq:replacement}
\ee
where ${\rm tan}\beta = \vev{H_u}/\vev{H_d}$ and ${\rm tan}\beta' = \vev{H_u'}/\vev{H_d'}$ are given by
\ba
\sin2\beta=\frac{2b}{m_{H_{u}}^{2}+m_{H_{d}}^{2}+2|\mu|^{2}},~~
\sin2\beta'=\frac{2b'}{m_{H'_{u}}^{2}+m_{H'_{d}}^{2}+2|\mu'|^{2}}.
\label{eq:sin2betabeta'}
\ea
The potential of $H$ and $H'$ is
\begin{align}
V(H,H')=&m^{2}(|H|^{2}+|H'|^{2})+\overline{\lambda}(|H|^{2}+|H'|^{2})^{2} \nonumber \\ &+\overline{\kappa}(|H|^{4}+|H'|^{4})\nonumber \\
&+\Delta m_{H}^{2}|H|^{2}+\overline{\rho}|H|^{4}\,,
\label{eq:THpotential}
\end{align}
where
\begin{align}
\bar{\lambda} =& \left(\frac{\lambda_X}{2}\right)^2 {\rm cos}2\beta {\rm cos}2\beta' , \\\label{eq:rho}
\bar{\rho}=&
\left[\left(\frac{\lambda_{X}}{2}\right)^{2}+\left(\frac{g_{2}^{2}+g_{Y}^{2}}{8}\right)\right](\cos^{2}2\beta-\cos^{2}2\beta')\,, \\\label{eq:kappa}
\bar{\kappa}=&\left(\frac{\lambda_{X}}{2}\right)^{2}\cos2\beta'(\cos2\beta'-\cos2\beta)+\left(\frac{g_{2}^{2}+g_{Y}^{2}}{8}\right)\cos^{2}2\beta'
\,, \\
m^2 =& |\mu'|^2 + m_{H_u'}^2 \sin^2 \beta' + m_{H_d'}^2 \cos^2 \beta' - b' \sin 2\beta' \,, \\
\label{eq:DeltamHsqure}
\Delta m_H^2 = &  \left(|\mu|^2 + m_{H_u}^2 \sin^2 \beta + m_{H_d}^2 \cos^2 \beta - b \sin 2\beta\right) \nonumber \\
& - \left(|\mu'|^2 + m_{H_u'}^2 \sin^2 \beta' + m_{H_d'}^2 \cos^2 \beta' - b' \sin 2\beta'\right) . 
\end{align}
In Eq.~(\ref{eq:THpotential}), the first line respects both $Z_{2}$ and the global $SU(4)$, the second line breaks $SU(4)$ explicitly while respecting $Z_{2}$, and the last line breaks $Z_{2}$ and $SU(4)$. $\Delta m_{H}^{2}|H|^{2}$ and $\overline{\rho}|H|^{4}$ are  soft and hard $Z_{2}$ breaking, respectively. 
In addition to these tree-level potentials, quantum corrections from the top Yukawa are important in determining the SM-like Higgs mass. We will discuss it in Sec.~\ref{sec:mHZ2}.

For $m^2 \gg \Delta m_H^2$ and $\bar{\lambda} \gg \bar{\kappa},\bar{\rho}$, the potential is approximately $SU(4)$ symmetric. 
The approximate $SU(4)$ symmetry is broken to $SU(3)$ by the condensation of $H'$, and the SM-like Higgs is understood as four pseudo-Nambu Goldstone bosons. (Three are eaten by $W'$ and $Z'$.) The EW scale is thus doubly protected by SUSY and the $SU(4)$ symmetry. See~\cite{Birkedal:2004xi,Chankowski:2004mq} for other models with double protection based on SUSY and a global symmetry.

The tree-level mass is enhanced in comparison with the MSSM. To see this, let us use the nonlinear realization of the Higgs $h$ as a pNGB of the broken $SU(4)$;
\be
H=f\sin\frac{h}{\sqrt{2}f}\quad,\quad H'=f\cos\frac{h}{\sqrt{2}f}\,,
\label{eq:nonlinear}
\ee
where $f^{2}=v^{2}+v^{'2}$ is the $SU(4)$-breaking scale.
For $\bar{\lambda}>\!\!>\bar{\kappa},\bar{\rho}$, the pNGB mass reads
\be
m_{h}^{2}\simeq4v^{2}(2\bar{\kappa}+\bar{\rho})\left(1-\frac{v^{2}}{f^{2}}\right)+\mathcal{O}(\bar{\kappa}/\bar{\lambda})\,,
\label{eq:mHsquare}
\ee
which is proportional to explicit $SU(4)$-breaking quartic couplings.
This is larger than the MSSM tree-level prediction $4 \bar{\kappa} v^2$, so a large stop mass is not required to explain the observed Higgs mass. In particular, for $f \gg v$, $\bar{\rho} \ll \bar{\kappa}$, and a large $\tan\beta$, the tree-level Higgs mass is already around 125 GeV.

The $Z_2$ breaking in the quadratic term $\Delta m_H^2$ is determined from $v$ and $v'$.  
In the limit $\bar{\kappa}>\!\!>\bar{\rho}$
and $f>\!\!>v$, combined with Eq.~(\ref{eq:mHsquare}), Eq.~(\ref{eq:DeltamHsqure}) reduces to
\ba
\Delta m_{H}^{2}\simeq\frac{1}{4}\left(\frac{f}{v}\right)^{2}m_{h}^{2}\simeq(200{\rm GeV})^{2}\times\left(\frac{f/v}{3}\right)^{2}\, 
  \simeq  \frac{0.07}{\bar{\lambda}}(- m^2) \equiv \Delta m^2_{H,{\rm req}}. 
\label{eq:mh2}
\ea

SUSY TH models may have two types of tuning: 1) the tuning of $\Delta m_{H}^{2}$ to obtain the hierarchy $v < f$ and 2) the tuning to obtain $f < m_{\rm SUSY}$.
The former is given by 
\be
\Delta_{v/f}= 
\frac{\partial\log(v^{2}/f^{2})}{\partial\Delta m_{H}^{2}}
 = \frac{f^{2}}{2v^{2}}-1\,.
\label{eq:FT1}
\ee
Avoiding the tuning worse than $10\%$ requires $f$ to be smaller than $5v$.
The latter is typically dominated by the quantum correction from the stop mass and is given by
\be
\Delta_{f}\equiv\frac{\log f^{2}}{\log m_{\tilde{t}}^{2}}\simeq\frac{3}{8\pi^{2}}\frac{y_{t}^{2}}{\bar{\lambda}}\frac{m_{\tilde{t}}^{2}}{f^{2}}\,.
\label{eq:FT2}
\ee
The total degree of fine-tuning is given by  the product $\Delta_{v/f}\times\Delta_{f}$,
\be
\Delta_{\rm TH}\equiv\Delta_{v/f}\times\Delta_{f}\simeq\frac{3}{8\pi^{2}}\frac{y_{t}^{2}}{\bar{\lambda}}\frac{m_{\tilde{t}}^{2}}{2v^{2}}\,.
\ee
In comparison with the fine-tuning in the MSSM, $(3 y_t^2 m_{\tilde{t}}^2/8\pi^2 \lambda_{\rm SM} v^2)$, the fine-tuning in SUSY TH models is improved by a factor of $2 \bar{\lambda}/\lambda_{\rm SM} \simeq 15 \bar{\lambda}$. Together with the large tree-level Higgs mass, which remove the necessity of a large stop mass, the EW scale can be obtained with a mild tuning of $O(1\mathchar`-10)$ \%.

%%%%%%%%%%%%%%%%%%%%%%%%%%%%%%%%%
\section{Soft Masses from $Z_2$-odd Polonyi Field}
\label{sec:model}
\setcounter{equation}{0}
%%%%%%%%%%%%%%%%%%%%%%%%%%%%%%%%%

We assume that the Polonyi field $S$ is odd under the $Z_2$ exchange symmetry in the TH mechanism. 
In this section, we show how the SM and mirror sector fields can couple to $S$ and obtain soft masses.

We assume that $S$ is the dominant source of supersymmetry breaking, so that its $F$-term is given by
\be
F_S = \sqrt{3} m_{3/2}M_P \equiv m_{3/2} M_*,
\ee
where $m_{3/2}$ is the gravitino mass and $M_{P}=2.4\times10^{18}{\rm GeV}$ is the reduced Planck mass. Here we define $M_* = \sqrt{3} M_P$ in order to remove the factors of $\sqrt{3}$ from the formulae for soft masses shown below.

We write down the coupling between $S$ and other fields in terms of dimension-less coefficients and $M_*$. If the cutoff scale $M_{\rm cut}$ is around $M_P$, we expect that dimension-less coefficients are ${\cal O}(1)$ .
When $M_{\rm cut}$ is below $M_P$, dimension-less coefficients are expected to be ${\cal O}\left(\left(M_P/M_{\rm cut}\right)^{D-2}\right)$ for the Kahler potential and ${\cal O}\left(\left(M_P/M_{\rm cut}\right)^{D-3}\right)$ for the superpotential and gauge kinetic terms, where $D$ is the dimension of the operator.

Since $F_S$ is $Z_2$ odd, the couplings of $S$ to SM and mirror fields seem to lead to large $Z_2$-breaking soft masses. However, a linear combination of the $Z_2$ symmetry and a $Z_{4R}$ subgroup of $R$ symmetry is unbroken by non-zero $F_S$, and the apparent $Z_2$ breaking can be removed by a discrete $R$ rotation.  Here we assign a vanishing $R$ charge to the scalar component of $S$. Physical $Z_2$-breaking arises only by picking-up $R$ symmetry breaking, which at the minimal is given by the non-zero VEV of the superpotential, namely, the gravitino mass.
The effect of the $R$ breaking can be expressed by the $F$ term of the conformal compensator $\phi$,
\be
\phi = 1 + F_\phi \theta^2,~~F_{\phi} = m_{3/2}.
\ee

As we will see, supergravity effects including anomaly mediation generate physical $Z_2$ breaking. 
A non-zero VEV of the scalar component of $S$ also breaks $R$ symmetry. However, as we will see in Sec.~\ref{sec:susybreaking}, it is much smaller than the Planck scale in supersymmetry breaking models without the Polonyi problem and introduces negligible $R$ symmetry breaking. In the following, we take $\langle S \rangle \ll M_P$ that will be justified in Sec.~\ref{sec:Polonyi}.

\subsection{Gaugino masses}
Usually, in gravity mediation, tree-level gaugino masses of order scalar masses require that the Polonyi field be singlet under any symmetry. This is because the kinetic terms of gauge multiplets are singlet. In supersymmetric TH mechanism, the kinetic terms transform into their $Z_2$ partner. Therefore, $S$ can couple to gauge multiplets via
\be
W\supset\,\, k_ig_{i}^{2}\frac{S}{M_{*}}{\rm Tr[}W^{\alpha}_{i}W_{\alpha i}-W^{'\alpha}_{i}W^{'}_{\alpha i}]\,,
\label{eq:gauginomass}
\ee
where
$g_i$ are gauge coupling constants,
$W^\alpha_i$ $(i=1,2,3)$ are the field strength superfields of the SM gauge fields with a spinor index $\alpha$, and $k_i$ are dimensionless coupling constants.

Eq.~(\ref{eq:gauginomass}) provides $Z_2$-odd gaugino masses and their values at the TeV scale are
$M_{3}\simeq1.3k_{3}m_{3/2}$, $M_{2}\simeq0.4k_{2}m_{3/2}$, and $M_{1}\simeq0.2k_{1}m_{3/2}$.
The anomaly mediation~\cite{Randall:1998uk,Giudice:1998xp} gives $Z_2$-even gaugino masses. Adding them together, we obtain
\be
M_{i}=-b_{i}\frac{\alpha_{i}}{4\pi}m_{3/2}+k_ig_{i}^{2}m_{3/2}\quad,\quad M_{i}'=-b_{i}^{'}\frac{\alpha^{'}_{i}}{4\pi}m_{3/2}-k_ig_{i}^{2}m_{3/2}\,,
\label{eq:gauginomass2}
\ee
where $b_{i}=b_{i}^{'}=(33/5,1,-3)$ and $\alpha_i\equiv g_i^{2}/(4\pi)$.%
\footnote{This is the first beta function coefficients in the MSSM, which need properly modified in the extensions including additional particle contents depending on models.}
As we anticipated, gaugino masses have physical $Z_2$ breaking only if both contributions are present.
When $M_{\rm cut} \sim M_P$, since $k_{i}=\mathcal{O}(1)$, the mass splitting between the MSSM and mirror gaugino is $O(0.1\mathchar`-1)$ \%, so gaugino mass terms remain nearly degenerate. When $M_{\rm cut} \ll M_P$, $k_{i}\sim M_{P}/M_{\rm cut} \gg 1 $ and even smaller splitting is expected. As we will see, this splitting can cause the splitting in the Higgs mass squared radiatively. After SM and mirror EW symmetry breaking, $v'>v$ introduces further mass splitting for the bino and wino.

Couplings of $S$ to charged fields $Q$ in the K\"{a}hler potential,
\be
\label{eq:SQQ}
K \supset \frac{S}{M_*} (Q^\dag Q - Q^{'\dag} Q^{'}),
\ee
also generate one-loop suppressed gaugino masses~\cite{DEramo:2012vvz}. These, however, are also $Z_2$-odd and negligible in comparison with the tree-level $Z_2$-odd one.

In $D$-term SUSY TH models, the new gauge multiplet whose $D$-term is responsible for the $SU(4)$ invariant quartic coupling should be $Z_2$ even so that it can couple to both the Higgs and Twin Higgs. The coupling of it to $S$ is forbidden by the $Z_2$ symmetry, and the gaugino mass is absent at tree-level. Potential large corrections to the soft masses of Higgses due to the large gauge coupling constant are automatically suppressed.

\subsection{$A$ terms}
$A$ terms are also obtained at tree-level. For example, the following superpotential
\be
W\supset d_u\frac{S}{M_{*}} y_u \left( Q H_{u} u^{c}- Q' H_{u}' u^{'c} \right),
\label{eq:yukawa}
\ee
with $Q$  ($u^{c}$) being $SU(2)_{L}$ doublet (singlet) and $Q'$  ($u^{'c}$) $SU(2)'_{L}$ doublet (singlet), gives rise to $Z_2$-odd $A$ terms. Here $y_u$ is the up-type yukawa coupling and $d_u$ is a dimensionless coupling constant. The K\"{a}hler potential of the form in \eqref{eq:SQQ} also gives $Z_2$-odd $A$ terms of the same order. On the other hand, the anomaly mediation~\cite{Randall:1998uk,Giudice:1998xp} gives $Z_2$-even $A$ terms. When combined, $A$-terms for the two sectors can be written as
\be
A^{u}=y_u d_u m_{3/2}-y_u (\gamma_{Q}+\gamma_{H_{u}}+\gamma_{u^{c}})m_{3/2}\equiv A^{u}_{S}+A^{u}_{\rm AMSB}\,,
\label{eq:Au1}
\ee
\be
A^{'u}=-y_u d_u m_{3/2}- y_u(\gamma_{Q'}+\gamma_{H'_{u}}+\gamma_{u^{'c}})m_{3/2}\equiv-A^{u}_{S}+A^{u}_{\rm AMSB}\,,
\label{eq:Au2}
\ee
where $\gamma_{X}$ is the anomalous dimension of the field $X$. When $M_{\rm cut} \sim M_P$, we expect $d_{u},d_{d}=\mathcal{O}(1)$ and the splitting in the $A$ terms is of $O(1)$\%. For $M_{\rm cut} \ll M_P$,
$d_{u},d_{d}\sim M_P/M_{\rm cut} \gg 1$ and
the splitting would be smaller than $\mathcal{O}(1)\%$.

\begin{figure}[!t]
\centering
	\includegraphics[width = 0.8\linewidth]{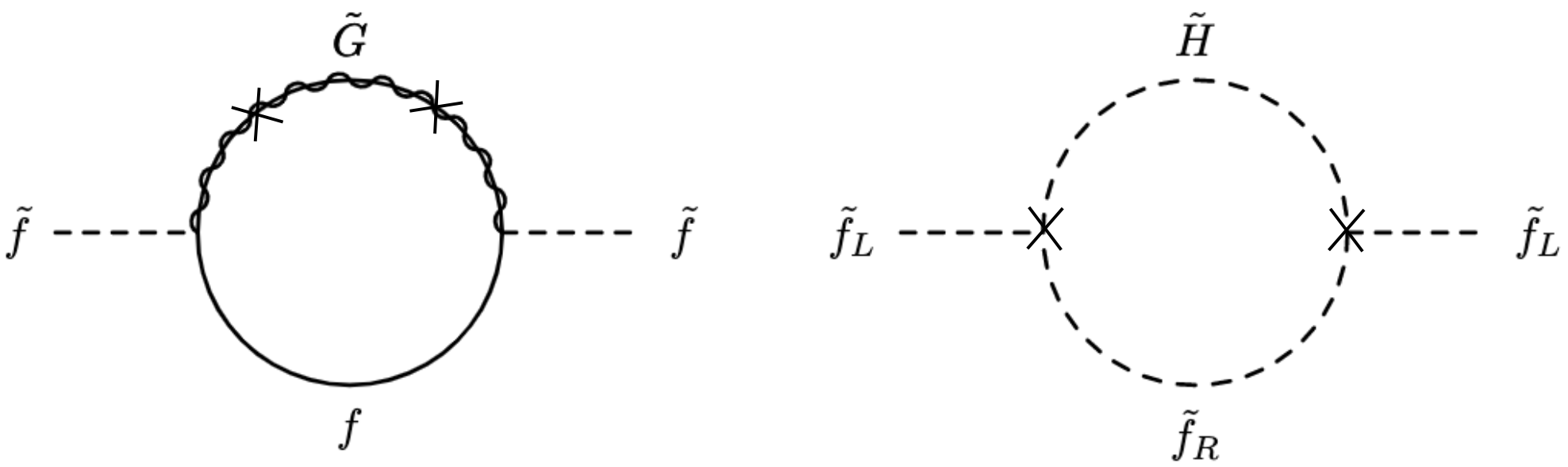} 
	\vspace*{-0mm}
	\caption{The one loop diagrams that result in difference in $\delta m_{\tilde{f}}$ and $\delta m_{\tilde{f}^{'}}$. The crosses represent gaugino masses (left panel) or scalar trilinear couplings (right panel).}
	\label{fig1}
\end{figure}

\subsection{Sfermion masses}
Sfermion masses $m_{\tilde{f}}$ are $O(m_{3/2} M_P/M_{\rm cut})$.
At tree-level, there is no $Z_2$-breaking soft masses, since $|F_S|^2$ is $Z_2$ even. $F_S F_\phi^\dag$ is $Z_2$-odd, but the K\"{a}hler potential $\phi^{\dagger} S Q Q^\dag$ is forbidden by the formal conformal symmetry at tree-level.
Such terms arise at loop-level and give $Z_2$-breaking soft masses. For example, quantum corrections from gaugino masses and/or $A$ terms generate $Z_2$-breaking soft scalar masses.

Since $m_{\tilde{f}}^{2}$ receives quantum corrections from gaugino masses and $A$ terms (see Appendix.~\ref{sec:appendixA}),
the splitting in these quantities between two sectors ($\Delta|M_{i}|^{2}$ and $\Delta|A^{f}|^{2}$) contributes to the RGE of the splitting in the sfermion mass $\Delta m_{\tilde{f}}^2\equiv m_{\tilde{f}}^{2}-m_{\tilde{f}'}^2$. Fig.~\ref{fig1} shows the one-loop diagrams contributing to these splittings. The interference between $Z_{2}$-even and odd contributions to gaugino masses and the trilinear couplings $A^{f}$ in Eqs.~(\ref{eq:gauginomass2}) and (\ref{eq:Au2}) respectively has the opposite sign for the two sectors and this leads to
\be
\Delta|M_{i}|^{2}=4\left(b_{i}\frac{\alpha_{i}}{4\pi}m_{3/2}\right)(k_{i}g_{i}^{2}m_{3/2})\quad,\quad\Delta|A^{f}|^{2}=4A^{f}_{S}A^{f}_{\rm AMSB}\,.
\label{eq:DeltaMA}
\ee

By substituting Eq.~(\ref{eq:DeltaMA}) into Eq.~(\ref{eq:RGEsfermion}) and solving RGEs, we estimate the sfermion mass splittings $\Delta m_{\tilde{f}}^{2}$ between the two sectors. We find that $\Delta m_{\tilde{f}}^{2}$ is dominated by $\Delta M_{i}^{2}$ and
\be
\Delta m_{\tilde{f}}^{2}\simeq\begin{cases}(8\times10^{-2})\times m_{3/2}\times M_{3}[1{\rm TeV}] & \text{for } \tilde{Q}_{L},\tilde{q}_{R},\\ -(8\times10^{-3})\times m_{3/2}\times M_{2}[1{\rm TeV}] & \text{for } \tilde{L}_{L},\\ -(5\times10^{-2})\times m_{3/2}\times M_{1}[1{\rm TeV}]& \text{for } \tilde{e}_{R},\end{cases}\,
\label{eq:mfsplitting}
\ee
where we write $\Delta m_{\tilde{f}}^{2}$ in terms of the gravitino mass and tree-level gaugino masses in Eq.~(\ref{eq:gauginomass}) evaluated at the TeV scale. The sign difference in the sfermion mass splitting for squarks and sleptons is ascribed to $b_{3}<0$ and 
$b_{2},b_{1}>0$.

%%%%%%%%%%%%%%%%%%%%%%%%%%%%%%%%%%%%%%%%%%%%
\subsection{Higgs mass parameters}
\label{sec:Vh}
%%%%%%%%%%%%%%%%%%%%%%%%%%%%%%%%%%%%%%%%%%%%
In this subsection, we study how the soft masses and the $\mu$-term of Higgses are generated.
We will see that the opposite signs of $S$ couplings to the SM and mirror operators give a $Z_{2}$-breaking Higgs potential. We then compute the $Z_2$-breaking mass in the potential of the SM-like Higgs and the twin Higgs after integrating out heavy Higgses.

\subsubsection{Tree-level mass terms}
\label{eq:treelevel}
We begin with the tree-level contributions to the $\mu$-term, $b$-term, and soft mass squared of the Higgses.
To be concrete, we consider $R[H_u H_d] =0$, so that the $\mu$ and $b$ terms can be naturally obtained from $R$ and supersymmetry breaking.%
\footnote{There can be another contribution to $\mu$ and $b$ terms from the direct coupling of the SUSY-breaking sector to $H_{u}H_{d}$ in the superpotential~\cite{Choi:2022fce,Choi:2022ssv}. As is shown in Appendix.~\ref{sec:appendixB}, appropriate size of $b$ terms may be obtained while $\mu$ terms cannot.}

First, we study the tree-level sources of  $\mu$-term and $b$-term from the K\"{a}hler potential. Given a K\"{a}hler potential $K$, one can obtain the leading contribution by expanding $-3e^{-K/(3M_{P}^{2})}$ below,
\be
\mathcal{L}\supset-3M_{P}^{2}\int d^{2}\theta d^{2}\bar{\theta}\,\,\phi^{\dagger}\phi e^{-K/(3M_{P}^{2})}\,.
\label{eq:Lsugra}
\ee
The following operators in K\"{a}hler potential including $H_{u}H_{d}$ are allowed by the $Z_2$ and $R$ symmetry,
\begin{align}
-3M_{P}^{2}e^{-K/(3M_{P}^{2})}\supset&c_{2}(H_{u}H_{d}+H'_{u}H'_{d})+\frac{r_{3}S^{\dagger} + r_{3*}S}{M_*}(H_{u}H_{d}-H'_{u}H'_{d}) \nonumber \\
+&r_{4}\frac{S^{\dagger}S}{M_*^{2}}(H_{u}H_{d} + H'_{u}H'_{d})+{\rm h.c.} +...\,,
\label{eq:GM1}
\end{align}
where $c$'s and $r$'s are the dimensionless coefficients of each operator and each subscript denotes the mass dimension of the associated operators. We use $c$ for operators consisting of $H_{u}$ and $H_{d}$ only whereas $r$'s are used for the couplings of $S$ to Higgses.

After scaling out $\phi$, using $\phi=1+F_{\phi}\theta^{2}$, and integrating over $\bar{\theta}^{2}$ on $\phi^\dag$ and $S^\dag$ and over $\theta^{2}$ on $H_u H_d$, we obtain $\mu$-parameters
\ba
\mu= c_{2}F_{\phi}+r_{3}\frac{F_{S}}{M_*}\,,~~
\mu'= c_{2}F_{\phi}-r_{3}\frac{F_{S}}{M_*}\,.
\label{eq:muparameter}
\ea
Integrating over $\theta^{2}\bar{\theta}^{2}$ on $\phi$, $\phi^{\dagger}$, $S$, and $S^{\dagger}$ gives $b$-parameters
\ba
b&=&-c_{2}F_{\phi}^{2}-r_{3}F_{\phi}\frac{F_{S}}{M_*}+r_{3*}F_{\phi}\frac{F_{S}}{M_*}+r_{4}\left(\frac{F_{S}}{M_*}\right)^{2}\,,\cr\cr
b'&=&-c_{2}F_{\phi}^{2}+r_{3}F_{\phi}\frac{F_{S}}{M_*}-r_{3*}F_{\phi}\frac{F_{S}}{M_*}+r_{4}\left(\frac{F_{S}}{M_*}\right)^{2}
\label{eq:bparameter}\,.
\ea
Note that the terms linearly proportional to $F_S$ are $Z_2$-odd. 
Using $F_{\phi}\simeq F_{S}/M_{*}\simeq m_{3/2}$, $\mu$ and $b$-parameters are given by
\ba
\mu&=&(c_{2}+r_{3})m_{3/2}\quad,\quad\mu'=(c_{2}-r_{3})m_{3/2}\,,\cr\cr
b&=&(-c_{2}-r_{3}+r_{3*}+r_{4})m_{3/2}^{2}\quad,\quad b'=(-c_{2}+r_{3}-r_{3*}+r_{4})m_{3/2}^{2}\,.
\label{eq:btree}
\ea
Physical $Z_2$ breaking arises from the coexistence of $Z_2$-even contributions proportional to $c_2$ and $r_4$, and $Z_2$-odd contributions proportional to $r_3$ and $r_{3*}$.

The tree-level Higgs soft mass squared comes from the following K\"{a}hler potential,
\ba
-3M_{P}^{2}e^{-K/(3M_{P}^{2})}&\supset&SS^{\dagger}+H_{u}H_{u}^{\dagger}+H'_{u}H^{'\dagger}_{u}
+r^{u}_{4}(\frac{S^{\dagger}S}{M_*^{2}}H_{u}H^{\dagger}_{u}+\frac{S^{\dagger}S}{M_*^{2}}H'_{u}H^{'\dagger}_{u})\,.
\label{eq:Hum}
\ea
Here we use the superscript $u$ for the dimensionless coefficients of couplings to the up-type Higgs. For the down-type Higgs, we use the superscript $d$.
Terms such as $S H_uH_u^\dag$ can be eliminated by $S$-dependent rotation of $H_u$ and $H_u^\dag$.
After scaling out $\phi$, one can see that there are no terms proportional to $S^\dag \phi^n H_u H_u^\dag$ without extra $S$. Therefore, a $Z_2$-odd soft mass term proportional to $F_S^{\dagger} F_\phi$ is suppressed by $\langle S\rangle / M_P$ and $m_{H_{u}}^{2}\simeq m_{H'_{u}}^{2}$ should hold at the tree-level. After integrating over $\theta^{2}\bar{\theta}^{2}$, we obtain 
\ba
m_{H_{u}}^{2} \simeq m_{H'_{u}}^{2} &\simeq &
r_{4}^{u}\left(\frac{F_{S}}{M_{*}}\right)^{2} 
\equiv r^{u}\left(\frac{F_{S}}{M_{*}}\right)^{2}\,.
\label{eq:Hum3}
\ea
The same applies for  $H_{d}$ and $H_{d}'$ as well with the coefficients $r^{u}$'s replaced with $r^{d}$'s;
\ba
m_{H_{u}}^{2}\simeq m_{H'_{u}}^{2}\simeq r^{u}m_{3/2}^{2}\,,~~ m_{H_{d}}^{2} \simeq m_{H'_{d}}^{2}\simeq r^{d}m_{3/2}^{2}\,.
\label{eq:softmasses}
\ea

Note that $\mu$ and $b$ terms generically have tree-level $Z_2$ breaking, which may induce too large $\Delta m_H^2$. We consider the following two cases where $\Delta m_H^2$ can be naturally suppressed:
\begin{itemize}
    \item Case I: $|r^{u}|\sim |r^{d}| = O(1)\gg c_{2}, r_{3},r_{3*},r_{4}$ at the TeV scale.
    \item Case II:
    $|r^{u}|\sim |r^{d}|\sim r_{4} \gg r_{3},r_{3*} \gg c_{2}$ at the TeV scale.
\end{itemize} 

Case I is naturally realized when $M_{\rm cut} \sim M_{P}$ and couplings involving $H_u H_d$ are suppressed, for example by an approximate symmetry under which $H_u H_d$ is charged. 
The $Z_{2}$-even and $Z_{2}$-odd contributions to $\mu$ and $b$ are comparable with each others, but since $\mu$ and $b$ are smaller than $m_{H_u}^2$ and $m_{H_d}^2$, $\Delta m_H^2$ is suppressed. For this, it suffices to have a hierarchy of parameters only around $0.3$, since $\Delta m_H^2$ has to be only smaller than ${\cal O}(0.1) m_H^2$ (see Eq.~(\ref{eq:mh2})) and is proportional to $\mu^2$ and $b^2$.

Case II is naturally realized when $M_{\rm cut} < M_{P}$.
We then expect $c_2 =O(1)$, $r_3,r_{3*} = O(M_{*}/M_{\rm cut})$, and $r_4,r_{u},r_d = O((M_{*}/M_{\rm cut})^{2})$ at the UV scale,
and the hierarchy $|r^{u}|\sim |r^{d}|\sim r_{4}>\!\!>r_{3},r_{3*}>\!\!>c_{2}$ is achieved.  $Z_{2}$-odd and even contributions to $\mu$ and $b$ dominate respectively and the $Z_2$ breaking $\Delta m_H^2$ is suppressed.
We also use re-scaled couplings $\tilde{r} = r \times  (M_{\rm cut}/ M_*)^{D_r-2}$, where $D_r$ is the dimension of the corresponding operators. For example, $\tilde{r}_4 = r_4 \times (M_{\rm cut}/ M_*)^{2} $. The natural values of the rescaled couplings are $O(1)$.

\begin{figure}[!t]
\centering
	\includegraphics[width = 0.8\linewidth]{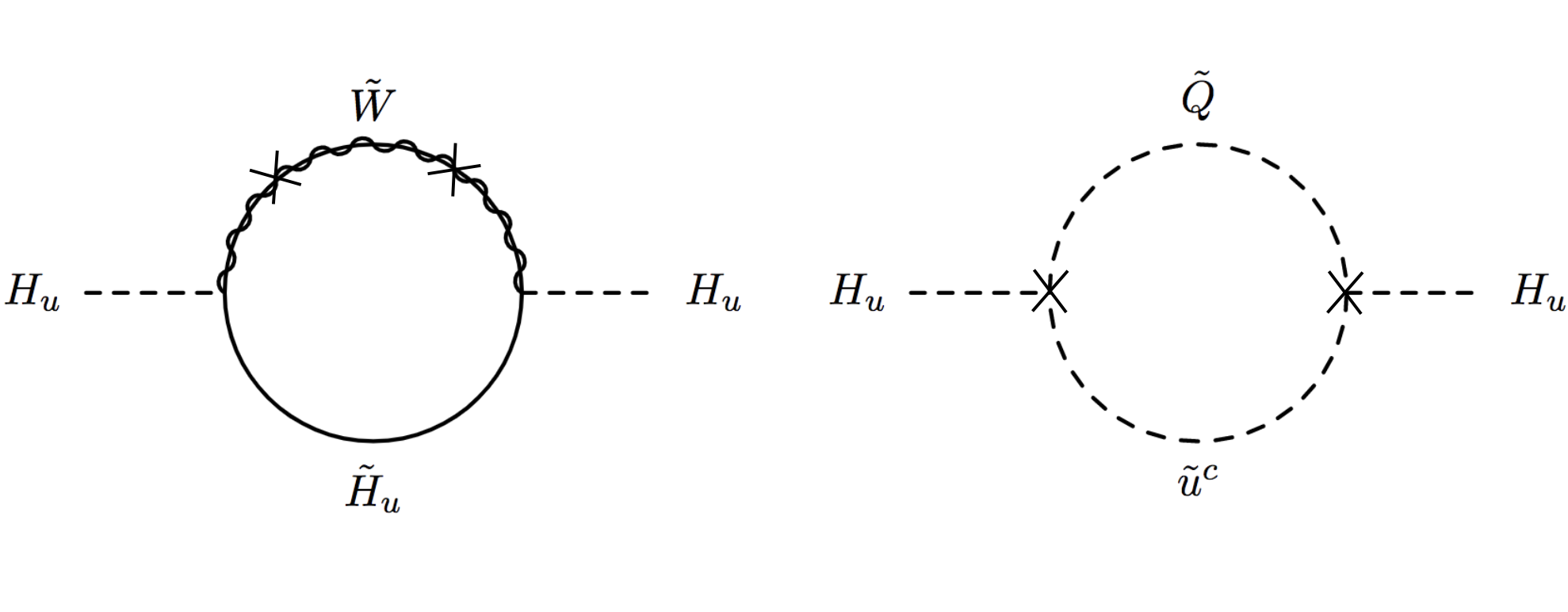} 
	\vspace*{-10mm}
	\caption{The one loop diagrams that result in difference in $\delta m_{H_{u}}^2$ and $\delta m_{H_{u}^{'}}^2$. The crosses represent the wino mass (left panel) or scalar trilinear couplings (right panel).}
	\label{fig2}
\end{figure}

\subsubsection{Quantum corrections}

Next, we investigate $Z_2$-breaking quantum corrections to the Higgs mass parameters. 
We first examine the one-loop radiative corrections $\delta m_{H_{u}}^2$ and $\delta m_{H_{d}}^2$ from the gauge and Yukawa interactions.
The splitting in wino masses in Eq.~\eqref{eq:gauginomass2} induces $\delta m_{H_{u}}^2-\delta m_{H'_{u}}^2\neq0$ from gauge interactions as shown in the left panel of Fig.~\ref{fig2}. The cross denotes the wino-mass insertion. Using the RGE for $m_{H_{u}}^{2}$ in Appendix.~\ref{sec:appendixA}, we obtain
\begin{align}
\Delta m_{H_{u},\,{\rm gauge}}^{2}&\equiv&\delta m_{H_{u}}^{2}-\delta m_{H'_{u}}^{2}\simeq\frac{ (4\times6)}{16\pi^{2}}\times\int_{\log[1{\rm TeV}]}^{\log[M_{\rm cut}]}dt g_{2}^{2}\left(b_{2}\frac{\alpha_{2}}{4\pi}m_{3/2}\right)(k_{2}g_{2}^{2}m_{3/2}) \nonumber \\
&=&(8\times10^{-3})\times m_{3/2}\times M_{2}[1{\rm TeV}]= (90~{\rm GeV})^2 \frac{m_{3/2}}{\rm TeV} \frac{M_{2}[1{\rm TeV}]}{\rm TeV},
\label{eq:masssplitting1}
\end{align}
where $t=d\log\mu$. The similar diagram for $H_{d}$ gives the similar $\Delta m_{H_{d},\,{\rm gauge}}^{2}\equiv\delta m_{H_{d}}^{2}-\delta m_{H'_{d}}^{2}$. 
$\Delta m_{H_{u},\,{\rm gauge}}^{2}$ and $\Delta m_{H_{d},\,{\rm gauge}}^{2}$ are smaller than $\Delta m^{2}_H$ in Eq.~(\ref{eq:mh2}) and do not introduce too much $Z_2$ breaking.

The splitting of the trilinear coupling in Eq.~(\ref{eq:yukawa}) induces
$\delta m_{H_{u}}^2-\delta m_{H'_{u}}^2 \neq0$ from the Yukawa interaction via the diagram shown in the right panel of Fig.~\ref{fig2}, where the cross denotes trilinear scalar coupling. Using the RGE for $m_{H_{u}}^{2}$ in Appendix.~\ref{sec:appendixA1} and the anomalous dimensions in Appendix.~\ref{sec:appendixA3}, we obtain
\ba
\Delta m_{H_{u},{\rm Yuk}}^{2}&\simeq&-\frac{(4\times6)}{16\pi^{2}}\times\int_{\log[1{\rm TeV}]}^{\log[M_{\rm cut}]}dt'\,\, A_{S}^{t}(t')A^{t}_{\rm AMSB}(t')\cr\cr
&\simeq&-\frac{(4\times6)}{16\pi^{2}}\times m_{3/2}\int_{\log[1{\rm TeV}]}^{\log[M_{\rm cut}]}dt'\,\, A_{S}^{t}(t')y_{t}(t')(\gamma_{Q_{3}}+\gamma_{H_{u}}+\gamma_{u^{c}_{3}}).
\label{eq:mHuyuk}
\ea
$A_{S}^{t}(t)$ in general depends on the UV boundary condition of it and the gaugino mass. To be concrete, we assume that $A_{S}^{t}(M_{\rm cut})=0$ and $dA_{S}^{t}/dt$ is dominantly determined by the one-loop correction from the top Yukawa, gauge interaction, and the gaugino mass; see Appendix.~\ref{sec:appendixA2}. 
We then obtain
\begin{align}
\Delta m_{H_{u},{\rm Yuk}}^{2} \simeq-0.01\times m_{3/2}\times M_{3}[1{\rm TeV}] 
\simeq -(200 {\rm GeV})^2 \frac{m_{3/2}}{\rm TeV} \frac{M_{3}[4{\rm TeV}]}{4 \rm TeV}.
\label{eq:masssplitting2}
\end{align}
$\Delta m_{H_{u},{\rm Yuk}}^{2} $ is typically larger than
$\Delta m_{H_{u},{\rm gauge}}^{2}$ and
may be as large as $\Delta m_H^2$ in Eq.~(\ref{eq:mh2}) to provide an appropriate size of $Z_2$ breaking.

There are radiative corrections to $b$ and $b'$ parameters as well.
The one-loop corrections $\delta b$ and $\delta b'$ are similar to what is shown in the left panel of Fig.~\ref{fig2} with the external line for $H_{u}^{\dagger}$ replaced with $H_{d}$ and a single mass insertion for each of the wino and Higgsino propagator instead of two mass insertions in the Higgsino propagator. 
If ${\rm sign}[b]= + (-){\rm sign}[b']$ at tree-level, quantum corrections such that $\delta b -\delta b' (\delta b + \delta b') \neq 0$ contribute to physical $Z_2$ breaking.
For Case II, the assumed hierarchy $r_{4}\gg r_{3},r_{3*},c_{2}$ ensures that ${\rm sign}[b]={\rm sign}[b']$.
For Case I, both signs are possible.
To be concrete, we discuss the case with ${\rm sign}[b]={\rm sign}[b']$ and compute $\delta b-\delta b'$.

In Case I, the correction comes from the tree-level $Z_2$-even $\mu$ term proportional to $c_2$ and the tree-level $Z_2$-odd wino mass.
Solving the RGE of $b$ and $b'$ in Eq.~(\ref{eq:RGEsAb}), we find $\delta b-\delta b'$ at the TeV scale to be 
\be
\Delta m_{H,b}^{2}\equiv \delta b-\delta b'\simeq-(400\,{\rm GeV})^{2}\frac{\mu[1{\rm TeV}]}{0.3{\rm TeV}}\frac{M_{2}[1{\rm TeV}]}{\rm TeV}\,,
\label{eq:deltab}
\ee
where we assume that the $\mu$ term is dominated by the $Z_2$-even contribution.
The contribution of $\Delta m_{H,b}^2$ to $\Delta m_H^2$ may be suppressed by $\sin 2\beta < 1$.
However, as we will see in Sec.~\ref{sec:SM-like higgs mass}, the consistency of the SM-like Higgs mass with a TeV scale stop mass requires $\tan\beta\sim 2-3$ and $\sin 2\beta \sim 0.7$ does not provide suppression. The resultant $\Delta m_H^2$ is comparable to the required one in \eqref{eq:mh2}. 

In Case II, although $\mu$ terms are dominated by the $Z_2$-odd terms, the dominant contribution to $\delta b-\delta b'$ comes from the interference between the tree-level $Z_{2}$-even contribution to $\mu$ proportional to $c_2$ and the tree-level $Z_{2}$-odd contribution to the gaugino mass. At the TeV scale it is
\be
\Delta m_{H,b}^{2}\equiv \delta b-\delta b'\simeq-(240{\rm GeV})^{2}\left(\frac{c_{2}}{0.1r_{3}}\right)\frac{\mu[1{\rm TeV}]}{{\rm TeV}}\frac{M_{2}[1{\rm TeV}]}{{\rm TeV}}\,,
\label{eq:deltab2}
\ee
which is as large as the required one in \eqref{eq:mh2} if $c_{2}\sim 0.1 r_{3}$, corresponding to $M_{\rm cut} \sim 0.1 M_*$.

The above quantum corrections to $b-b'$ may be understood as corrections to $r_3-r_{3*}$ and suggest that the assumed hierarchy for Case I and II is natural. Other types of hierarchy are not radiatively stable. 
For example, the hierarchy $|r^u| \sim |r^d| \sim c_2 = O(1) \gg r_{3},r_{3*},r_4$ could suppress the $Z_2$-breaking $\mu$ and $b$ terms, but the correction in Eq.~\eqref{eq:deltab} makes this hierarchy radiatively unstable.

%%%%%%%%%%%%%%%%%%%%%%%%%%%%%%%%%%%%%%%%%%%%
\section{Higgs potential and Higgs Mass}
\label{sec:mHZ2}
%%%%%%%%%%%%%%%%%%%%%%%%%%%%%%%%%%%%%%%%%%%%

In this section, we compute the potential of the Higgs and Twin Higgs in the decoupling limit and compute the mass of the SM-like Higgs.

\subsection{The difference of $\tan \beta$}

In this subsection, we compute the difference between $\tan \beta$ and $\tan \beta'$ and its effect on the Higgs mass terms in the decoupling limit.

\subsubsection*{Case I}

For the Higgs potential to be bounded from below, $|2b'|<m_{H_{u}'}^{2}+m_{H_{d}'}^{2}+2|\mu'|^{2}$ is required. In terms of the parameters introduced in Sec.~\ref{eq:treelevel}, 
\be
2|c_{2}-r_4 - r_3 + r_{3*}|<2(c_{2}-r_3)^{2}+r^{u}+r^{d}\,.
\label{eq:EWSB}
\ee
Without loss of generality, we assume $c_2 - r_4 >0$. We also assume $|r_3 - r_{3*}| < c_2 - r_4$ for simplicity. $b,b'<0$ for this case, so we flip the sign of $H_uH_d$ to take $b,b'>0$. 
Then from Eq.~(\ref{eq:sin2betabeta'}), $\sin2\beta$ and $\sin2\beta'$ are given by 
\be
\sin2\beta=\frac{2(c_{2}-r_4)+2(r_3-r_{3*})}{2(c_{2}+r_3)^{2}+r^{u}+r^{d}},\quad\sin2\beta'=\frac{2(c_{2}-r_4)-2(r_3-r_{3*})}{2(c_{2}-r_3)^{2}+r^{u}+r^{d}}\,.
\label{eq:sin2beta}
\ee
$Z_2$ breaking in the Higgs potential gives a small difference
between ${\rm tan}\beta$ and ${\rm tan}\beta'$.
Using the relation $\tan\beta\simeq(2/\sin2\beta)-(\sin2\beta/2)$ and denoting the leading quantum corrections to $\mu$, $b$ and soft masses squared of Higgs as $\delta\mu$, $\delta b$, $\delta m_{H_{u}}^{2}$ and $\delta m_{H_{d}}^{2}$, we find $\Delta\tan\beta$ at leading order is given by
\ba
\Delta\tan\beta&\simeq&[(m_{H_{u}}^{2}+m_{H_{d}}^{2})(b'-b+\delta b'-\delta b)+2\mu^{2}\delta b'-2\mu'^{2}\delta b\cr\cr
&&+b'\{\delta m_{H_{u}}^{2}+\delta m_{H_{d}}^{2}+2(\mu^{2}+\delta\mu^{2})\}-b\{\delta m_{H'_{u}}^{2}+\delta m_{H'_{d}}^{2}+2(\mu'^{2}+\delta\mu'^{2})\}]\cr\cr
&&\times(bb'+b'\delta b+b\delta b')^{-1}\,.
\label{eq:Deltat1}
\ea
Taking into account the assumed hierarchy among parameters, 
this can be approximated as
\begin{align}
\Delta\tan\beta \approx & \frac{(r^{u}+r^{d})[2(r_{3*}-r_{3})-0.3k_{2}(c_{2}+r_{3})]-0.6k_{2}(c_{2}+r_3)(c_{2}^{2}+r_{3}^{2})}{(c_{2}-r_{4})^{2}-(r_{3}-r_{3*})^{2}} \nonumber \\
+ & \frac{(c_{2}-r_{4})(\mathcal{O}(10^{-2})k_{3}+8c_{2}r_{3})-(r_{3}-r_{3*})[4(c_{2}^{2}+r_{3}^{2})]}{(c_{2}-r_{4})^{2}-(r_{3}-r_{3*})^{2}}\nonumber \\
\approx &  \frac{(r^{u}+r^{d})(2(r_{3*}-r_{3})-0.3k_{2}(c_{2}+r_{3}))+8(c_{2}-r_{4})c_{2}r_{3}}{(c_{2}-r_{4})^{2}}\,,
\label{eq:approxDeltat22}
\end{align}
where we used $|r_3 - r_{3*}| < |c_2-r_4|$ in the second equality.

Let us assess the effect of $\Delta {\rm tan}\beta$ on $Z_2$ breaking in the Higgs mass terms.
In the decoupling limit, the quadratic terms of the Higgs and Twin Higgs are
\begin{align}
V(H^,H')_{\rm quadratic}
&=\left[(c_{2}+r_{3})^{2}+r^{d}\cos^2\beta +r^{u}\sin^{2}\beta-(c_{2}-r_{4}+r_{3}-r_{3*})\sin2\beta\right]m_{3/2}^{2}|H|^{2} \nonumber \\
&+\left[(c_{2}-r_{3})^{2}+r^{d}\cos^2\beta' +r^{u}\sin^{2}\beta'-(c_{2}-r_{4}-r_{3}+r_{3*})\sin2\beta'\right]m_{3/2}^{2}|H'|^{2}\,. \nonumber \\
\label{eq:VHH}
\end{align}
The $Z_2$-breaking mass of the Higgs from $\Delta \tan \beta$ is, to the leading order in $(\Delta\tan\beta)/\tan\beta$,
\begin{align}
\Delta m_{H,\Delta\beta}^{2}\simeq&
\left(r^u - r^d + (c_2-r_{4}  )\tan\beta\right) \frac{2 \Delta\tan\beta}{\tan^3 \beta} m_{3/2}^2 \\
\simeq&\frac{r^u - r^d + (c_2-r_{4}) \tan\beta}{(c_{2}-r_{3})^{2}+r^{d}\cos^2\beta' +r^{u}\sin^{2}\beta'-(c_{2}-r_{4})\sin2\beta'} \frac{2 \Delta\tan\beta}{\tan^3 \beta}m_{H'}^{2}\,. \nonumber  
\label{eq:soft1}
\end{align}
As we will see in Sec.~\ref{sec:SM-like higgs mass}, the observed Higgs mass and the stop mass of ${\cal O}(1)$ TeV requires that ${\rm tan}\beta = 2\mathchar`-3$. $\Delta\tan\beta$ should not be larger than $O(1)$ in order to suppress $\Delta m_{H,\Delta \beta}^{2}$.
For $c_2,r_4 < r^{u,d}$, to suppress the term proportional to $r_{3*}-r_3$ in Eq.~\eqref{eq:approxDeltat22} requires that $r_{3*}-r_3$ be somewhat smaller than $c_2-r_4$.

\subsubsection*{Case II}

The analysis for Case II is parallel to that of Case I. Without loss of generality, we take $r_4 >0$. The Higgs potential is bounded from below if
\be
2r_{4}<2r_{3}^{2}+r^{u}+r^{d}\,.
\label{eq:EWSB2}
\ee
$\sin2\beta$ and $\sin2\beta'$ at the leading order are 
\be
\sin2\beta,\,\,\sin2\beta'\simeq\frac{2r_{4}}{r^{u}+r^{d}+2r_{3}^{2}}=\frac{2\tilde{r}_{4}}{\tilde{r}^{u}+\tilde{r}^{d}+2\tilde{r}_{3}^{2}}\,,
\label{eq:sin2beta2}
\ee
where we used $r^{u}=\tilde{r}^{u}(M_{*}/M_{\rm cut})^{2}$, $r^{d}=\tilde{r}^{d}(M_{*}/M_{\rm cut})^{2}$, $r_{3}=\tilde{r}_{3}(M_{*}/M_{\rm cut})$ and $r_{4}=\tilde{r}_{4}(M_{*}/M_{\rm cut})^{2}$ for the last equality.

$\Delta\tan\beta$ is given by the generic formula in Eq.~(\ref{eq:Deltat1}).
Taking into account the hierarchy of the parameters,
\begin{align}
\Delta\tan\beta &\approx 
\frac{2(r^{u}+r^{d})(r_{3}-r_{3*})-(4\times10^{-3})r_{3}^{3}-(c_{2}-r_{4})(\mathcal{O}(10^{-2})k_{3}+8c
_{2}r_{3})}{r_{4}^{2}}
\nonumber \\
 &\approx  \mathcal{O}(10^{-2})\frac{k_{3}}{r_{4}} + \frac{2(r^{u}+r^{d})(r_{3}-r_{3*})}{r_{4}^{2}} \nonumber \\
 &= \mathcal{O}(10^{-2}) \frac{M_{\rm cut}}{M_*}\frac{\tilde{k}_{3}}{\tilde{r}_{4}} + \frac{M_{\rm cut}}{M_*}\frac{2(\tilde{r}^{u}+\tilde{r}^{d})(\tilde{r}_{3}-\tilde{r}_{3*})}{\tilde{r} _{4}^{2}}
 \label{eq:approxDeltat222}
\end{align}
For ${\cal O}(1)$ couplings with tildes, the second term dominates and $\Delta\tan\beta=\mathcal{O}(M_{\rm cut}/M_{*}) \ll 1$.

In the decoupling limit, the quadratic terms of the Higgs and Twin Higgs are
\begin{align}
V(H,H')_{\rm quadratic}
&=\left[r_{3}^{2}+r^{d}\cos^2\beta +r^{u}\sin^{2}\beta-r_{4}\sin2\beta\right]m_{3/2}^{2}|H|^{2} \nonumber \\
&+\left[r_{3}^{2}+r^{d}\cos^2\beta' +r^{u}\sin^{2}\beta'-r_{4}\sin2\beta'\right]m_{3/2}^{2}|H'|^{2}\,,
\label{eq:VHH2}
\end{align}
where we have included only the $Z_2$ breaking through $\beta \neq \beta'$.
$Z_2$-breaking mass is, to the leading order in $(\Delta\tan\beta)/\tan\beta$,
\ba
\Delta m_{H,\Delta\beta}^2&\simeq&
(r^u - r^d + r_4 \tan\beta) \frac{2 \Delta\tan\beta}{\tan^3 \beta} m_{3/2}^2
\cr\cr
&\simeq&\frac{r^u - r^d + r_4 \tan\beta}{r^{d}\cos^2\beta' +r^{u}\sin^{2}\beta'-r_{4}\sin2\beta'} \frac{2 \Delta\tan\beta}{\tan^3 \beta}m_{H'}^{2}\,. 
\label{eq:soft2}
\ea
For $|r^{u}|\sim |r^{d}|\sim r_{4}$ and ${\rm tan}\beta = 2\mathchar`-3$,
$\Delta m_{H\Delta \beta}^2$ may be as large as $\Delta m_{H,{\rm req}}^2$ in Eq.~(\ref{eq:mh2}) if $\Delta\tan\beta=\mathcal{O}(M_{\rm cut}/M_{*})$ is not too much smaller than 1.

\subsection{The SM-like Higgs mass}
\label{sec:SM-like higgs mass}

In this subsection, we compute the SM-like Higgs mass and determine the preferred value of the stop mass and $\tan\beta$.

The Higgs potential is given by Eq.~(\ref{eq:THpotential}) plus quantum corrections around the soft mass scale, which are
dominated by the top Yukawa coupling. The one loop correction is given by
\begin{align}
V_{\cancel{SU(4)}}^{\rm top}\simeq&\frac{3}{16\pi^{2}}(\hat{g}_{t}^{2}|H|^{2}+m_{\tilde{t}}^{2})^{2}\left[\log\left(\frac{\hat{g}_{t}^{2}|H|^{2}+m^{2}_{\tilde{t}}}{\mu_{R}^{2}}\right)-\frac{3}{2}\right]-\frac{3}{16\pi^{2}}\hat{g}_{t}^{4}|H|^{4}\left[\log\left(\frac{\hat{g}_{t}^{2}|H|^{2}}{\mu_{R}^{2}}\right)-\frac{3}{2}\right] \nonumber \\
+&\frac{3}{16\pi^{2}}(\hat{g}'^{2}_{t}|H'|^{2}+m_{\tilde{t}}^{2})^{2}\left[\log\left(\frac{\hat{g}'^{2}_{t}|H'|^{2}+m^{2}_{\tilde{t}}}{\mu_{R}^{2}}\right)-\frac{3}{2}\right]-\frac{3}{16\pi^{2}}\hat{g}'^{4}_{t}|H'|^{4}\left[\log\left(\frac{\hat{g}'^{2}_{t}|H'|^{2}}{\mu_{R}^{2}}\right)-\frac{3}{2}\right]\,,\nonumber\\
\label{eq:VSU(4)breaking2}
\end{align}
where $\mu_{R}$ is a renormalization scale, $\hat{g}_{t}\equiv y_{t}$, and $\hat{g}'_{t}\equiv y_{t}\sin\beta'/\sin\beta$ with $y_{t}$ the top Yukawa coupling.
We assumed identical left-handed and right-handed stop masses $m_{\tilde{t}}$ for simplicity.

\begin{figure*}[!t]
  \centering
  \hspace*{-1mm}
\subfigure{\includegraphics[scale=0.6]{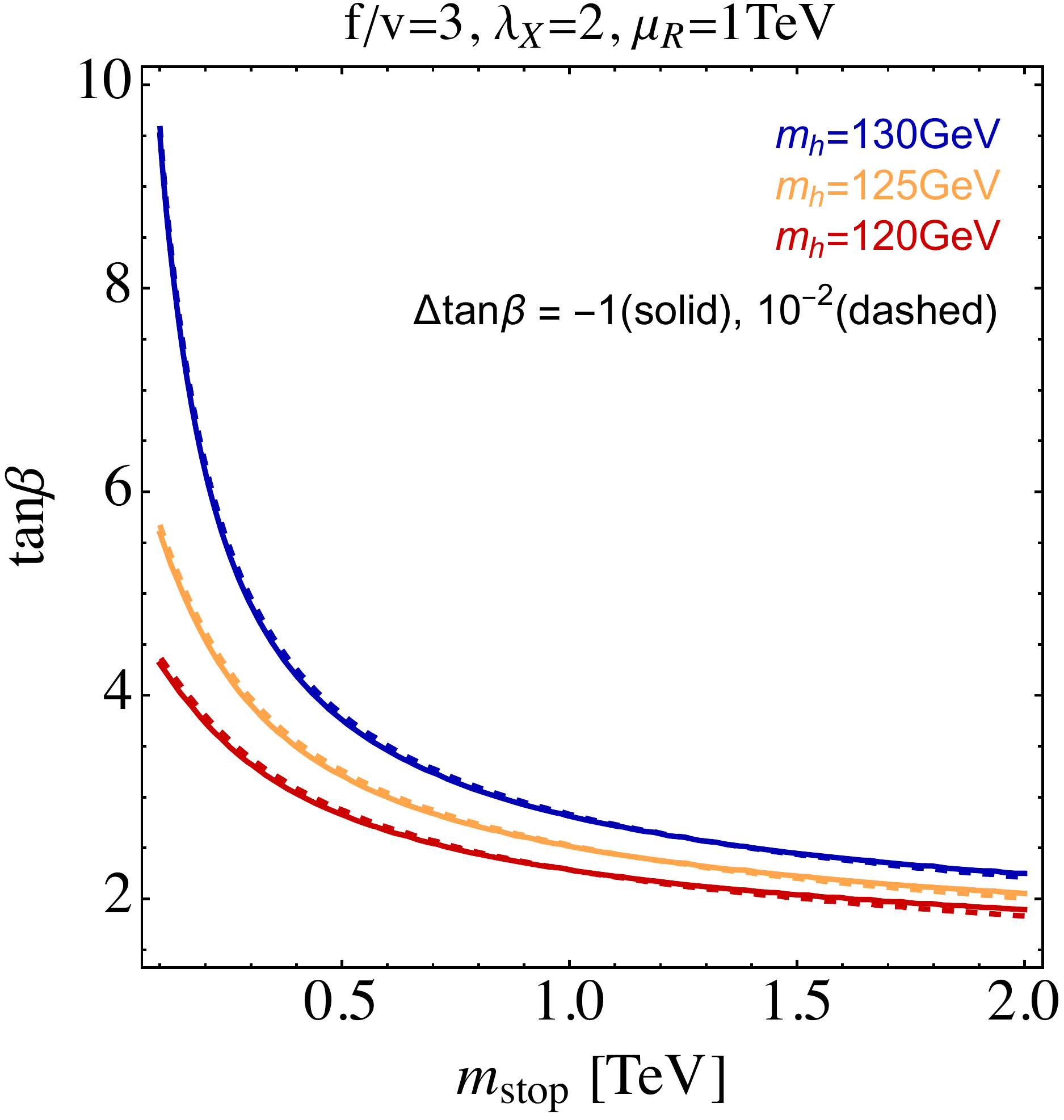}}
  \caption{The SM-like Higgs mass as a function of $m_{\rm stop}$ and $\tan\beta$. For the solid and dashed lines, $\Delta\tan\beta\equiv\tan\beta-\tan\beta'=-1$ and $10^{-2}$ are assumed, respectively.}
  \vspace*{-1.5mm}
\label{fig3}
\end{figure*}

In Fig.~\ref{fig3}, we show the contours of the SM-like Higgs mass computed from the potentials in Eqs.~(\ref{eq:THpotential}) and (\ref{eq:VSU(4)breaking2}). In the computation, we take $v=174{\rm GeV}$, $f/v=3$, $\lambda_{X}=2$, and $\mu_{R}=1{\rm TeV}$. Each solid and dashed line corresponds to $\Delta\tan\beta=-1$ and $10^{-2}$ respectively. We find that the predicted Higgs mass is insensitive to  $\Delta\tan\beta$ as long as $\Delta\tan\beta\lesssim1$. 
The Higgs mass decreases as $f/v$ or $\lambda_{X}$ decrease.  To satisfy the experimental lower bound on the stop mass of 1 TeV~\cite{ATLAS:2020dsf,ATLAS:2020xzu,CMS:2021eha}, $\tan\beta\lesssim3$ is required.

\begin{figure*}[htp]
  \centering
  \hspace*{-1mm}
  \subfigure{\includegraphics[scale=0.425]{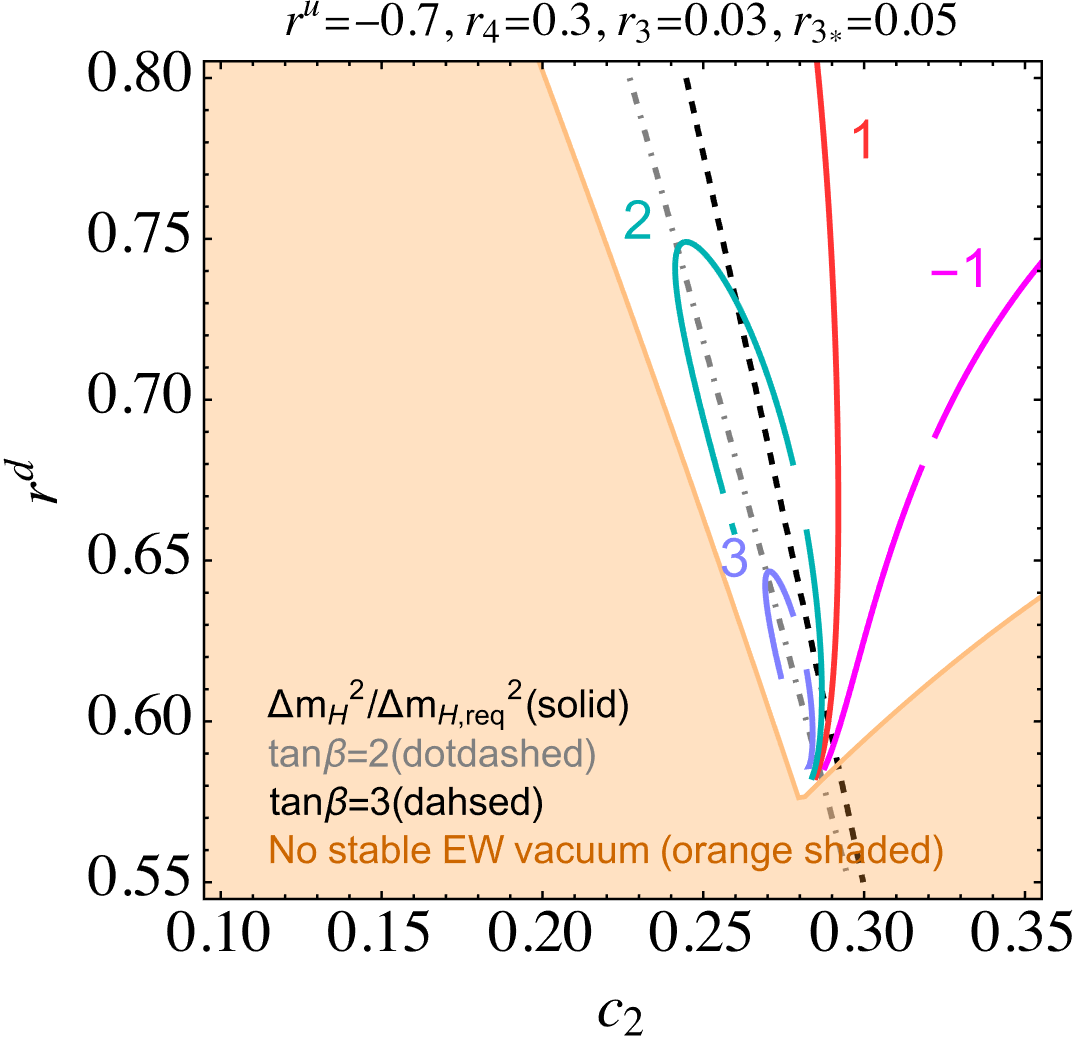}}
\subfigure{\includegraphics[scale=0.425]{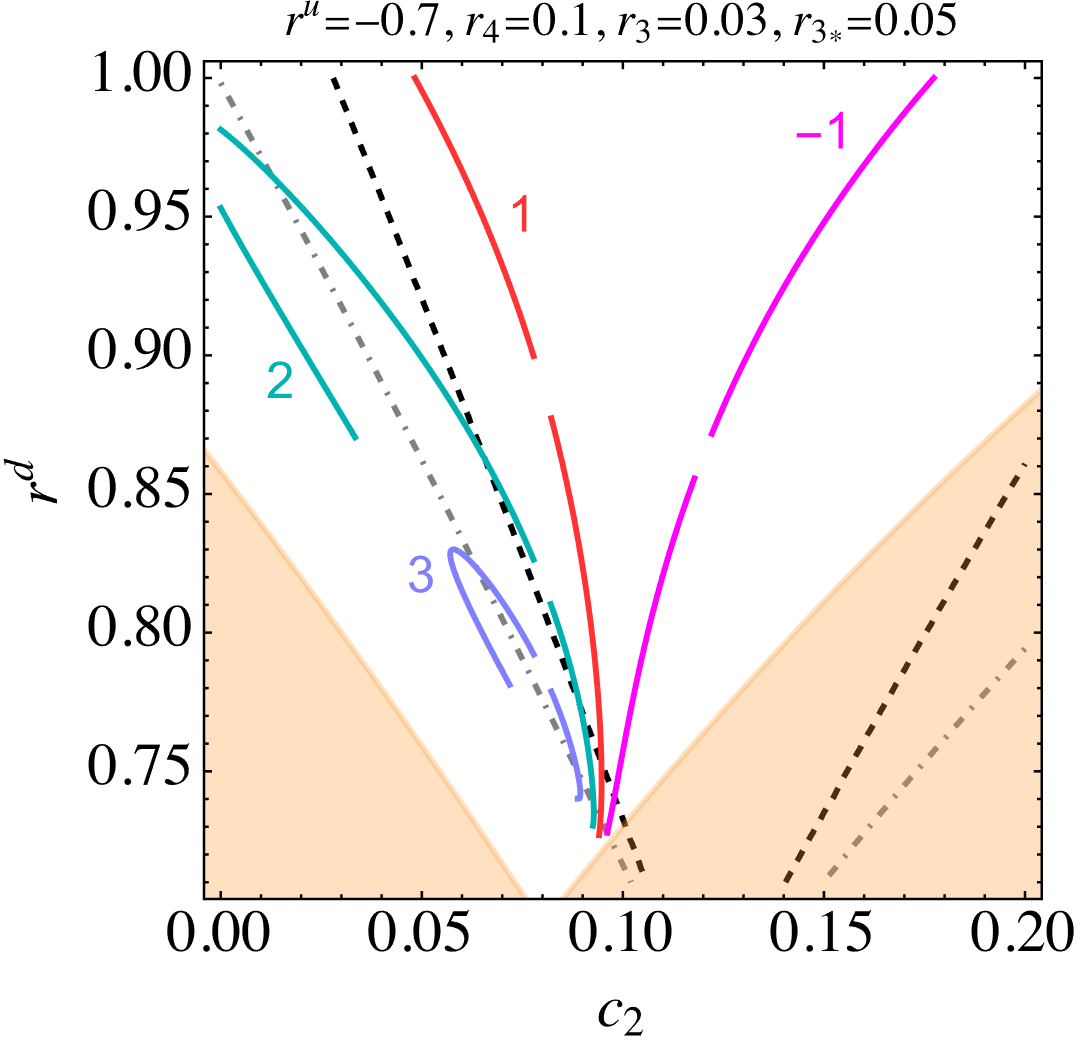}}
\subfigure{\includegraphics[scale=0.425]{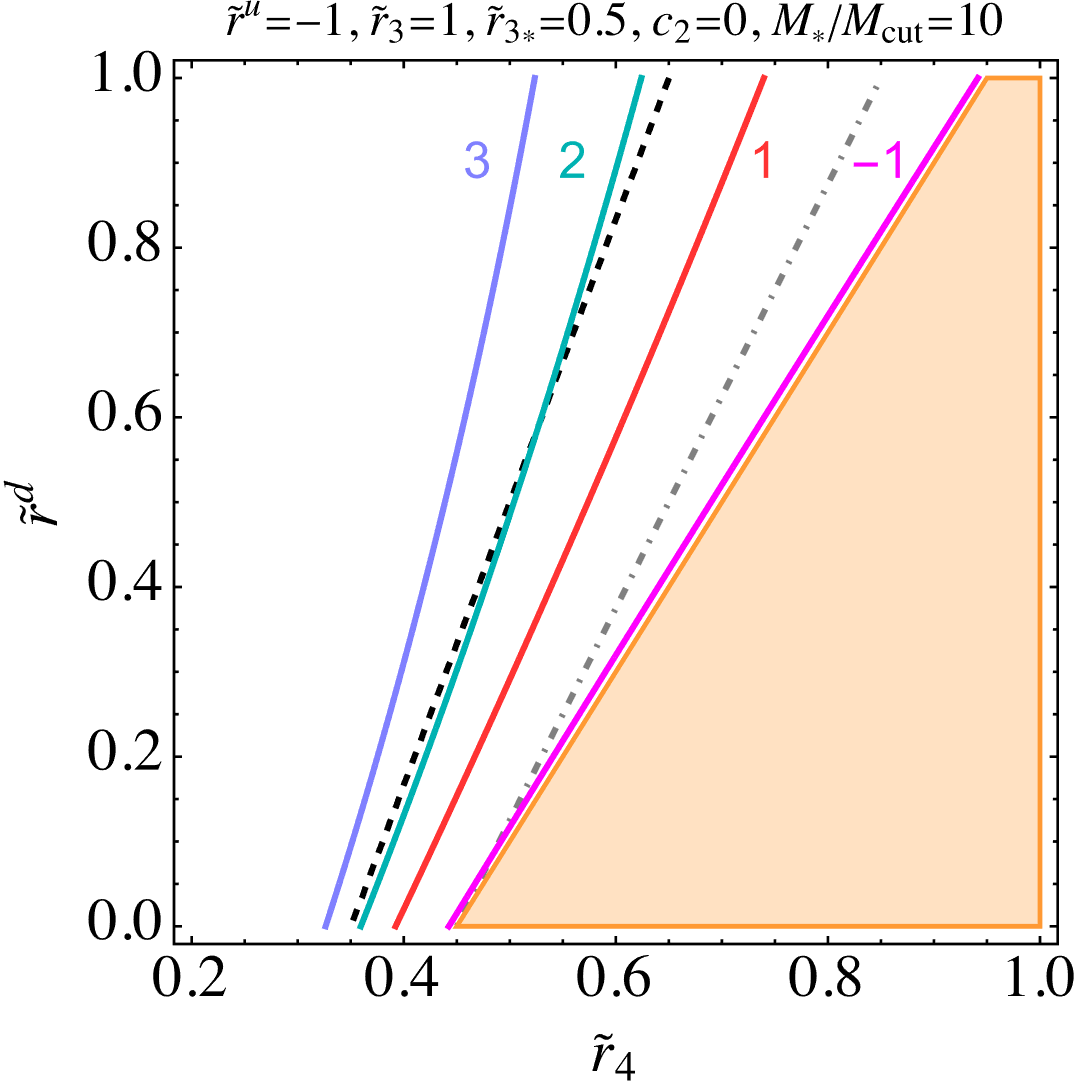}}
\subfigure{\includegraphics[scale=0.425]{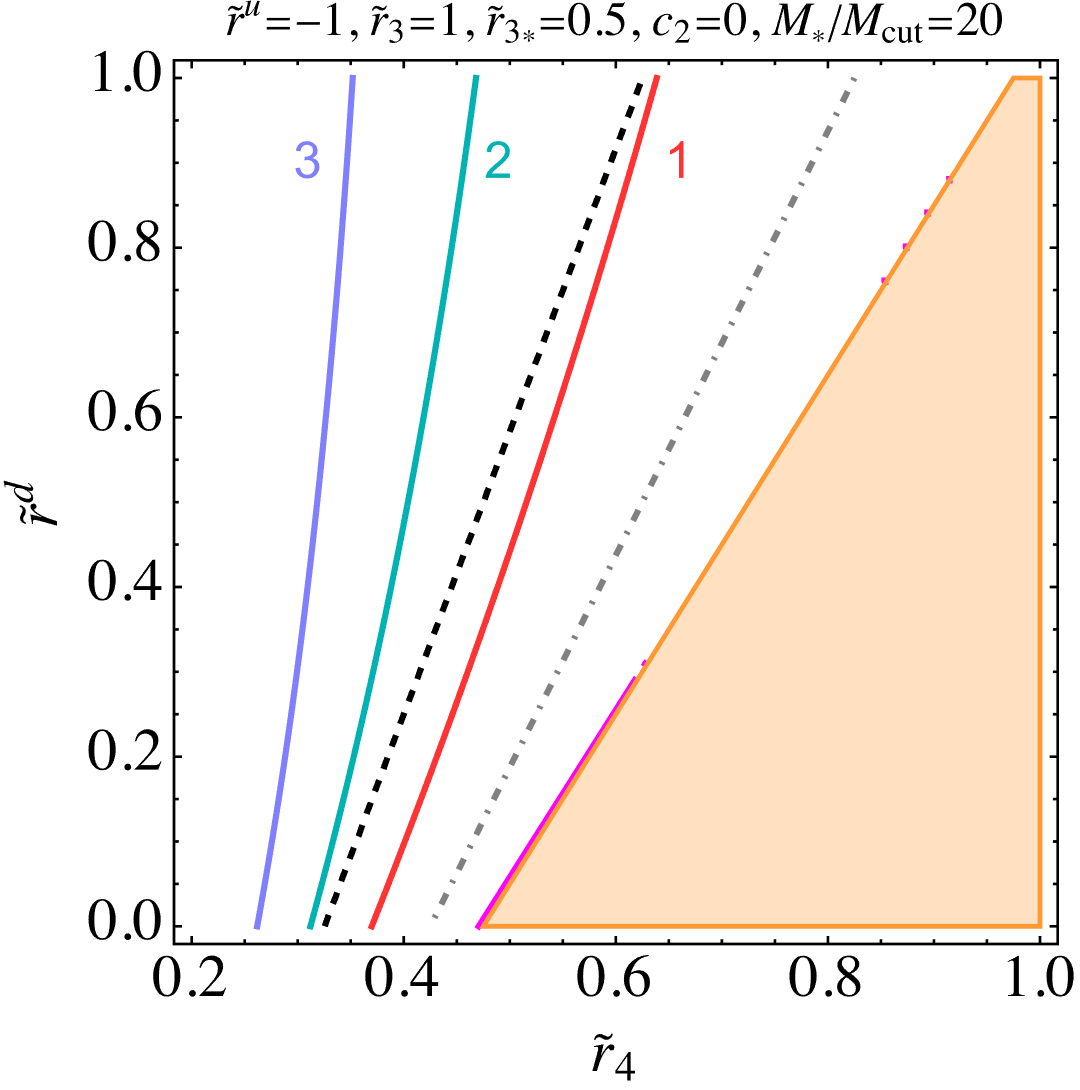}}
  \caption{
  Contours of $\tan \beta$ and the tree-level $Z_2$-breaking Higgs mass term $\Delta m_H^2$. There is no stable EW vacuum in the orange shaded region. Upper and lower panels correspond to Case I and II, respectively.}
  \vspace*{-1.5mm}
\label{fig4}
\end{figure*}

In Fig.~\ref{fig4}, we show 
the contours of $\tan\beta$ and the tree-level $\Delta m_{H}^{2}$ in the $(c_{2},r^{d})$ and $(\tilde{r}_{4},\tilde{r}^{d})$ planes.
Here we normalize $\Delta m_{H}^{2}$ by the required value $\Delta m_{H,{\rm req}}^{2}$ in Eq.~(\ref{eq:mh2}).
The upper and lower panels correspond to case I and II, respectively. In the orange shaded regions, stable mirror EW symmetry-breaking vacua  do not exist, i.e., $2 b' > m_{H'_{u}}^{2} + m_{H'_{d}}^{2} + 2 |\mu'|^{2} $ or $b'^{2}<(|\mu'|^{2}+m_{H'_{u}}^{2})(|\mu'|^{2}+m_{H'_{d}}^{2})$.

For Case I, as an example, we took $r^{u}=-0.7$. The gray dot-dashed and black dashed lines give $\tan\beta=2$ and $3$, respectively.
For $r_{3},r_{3*}<c_{2},r_{4}=\mathcal{O}(0.1)<r^{u,d}=\mathcal{O}(1)$, $|\Delta\tan\beta|=\mathcal{O}(0.1)$ according to Eq.~(\ref{eq:approxDeltat22}).
For the TeV scale  $m_{\rm stop}$, the observed Higgs mass is explained for $\tan\beta=2-3$,
which can be achieved when $m_{H_{u}}^{2}+m_{H_{d}}^{2}\sim b$, i.e., $r^{u}+r^{d}\sim c_{2}-r_{4}$. Along with $c_{2}=0.1-0.3$ for suppressing $\Delta m_{H,b}^{2}$ in Eq.~(\ref{eq:deltab}), this requirement results in the parameter space shown in two upper panels in Fig.~\ref{fig4}. The comparison of the two panels shows that a smaller $r_{4}$ requires a smaller $c_{2}$ and more fine-tuned cancellation between $r^{u}$ and $r^{d}$ for maintaining $\tan\beta\sim3$ and the sufficient suppression of $\Delta\tan\beta$.

For Case II, as an example, we take ($\tilde{r}^{u},\tilde{r}_{3},\tilde{r}_{3*},c_{2},M_{*}/M_{\rm cut})=(-1,1,0.5,0,10)$ and $(-1,1,0.5,0,20)$.
As $c_{2}$ increases, the orange shaded region becomes broader and the viable parameter space becomes narrower.
On the other hand, as $M_{*}/M_{\rm cut}$ increases, the $Z_2$ breaking in the Higgs mass terms becomes smaller and the contours of $\Delta m_H^2$ moves to the left, so that the tree-level $Z_2$ breaking cannot provide the required $\Delta m_H^2$ starting from $M_{*}/M_{\rm cut}\sim40$.
Also, for a given soft mass scale, as $M_{*}/M_{\rm cut}$ increases, $m_{3/2}$ becomes smaller, and $\Delta m_H^2$ from quantum corrections becomes smaller.
Therefore, obtaining the required amount of $\Delta m_H^2$ from tree-level or one-loop level $Z_2$ breaking requires that $m_{3/2}>\mathcal{O}(100){\rm GeV}$.

%%%%%%%%%%%%%%%%%%%%%%%%%%%%%%%%%%%%%%%%%%%%
\section{UV Completion and Early Universe Cosmology of Polonyi Field}
\label{sec:Polonyi}
%%%%%%%%%%%%%%%%%%%%%%%%%%%%%%%%%%%%%%%%%%%%

In this section, we present a concrete UV completion where the  $Z_2$-odd Polonyi field obtains a large enough mass around the origin. We then discuss the cosmological evolution of the Polonyi field.

%%%%%%%%%%%%%%%%%%%%%%%%%%%%%%%%%%%%%%%%%%%%
\subsection{SUSY-breaking sector}
\label{sec:susybreaking}
\setcounter{equation}{0}
%%%%%%%%%%%%%%%%%%%%%%%%%%%%%%%%%%%%%%%%%%%%

We construct Planck scale mediated SUSY-breaking scenarios where the SUSY-breaking occurs in a hidden sector. To obtain gaugino masses as large as scalar masses, the SUSY-breaking field should be a gauge-singlet.
To be concrete, we analyze the Izawa-Yanagida-Intriligator-Thomas model~\cite{Izawa:1996pk,Intriligator:1996pu}, but similar discussion is applicable to generic O’Raifeartaigh-type SUSY-breaking models with gauge-singlet SUSY-breaking fields.

\begin{table}[t]
\centering
\begin{tabular}{|c||c|c|c|c|c|c|} \hline
 & $Q_{i}$ & $S_{ij}$ & $H_{u}$& $H_{u}'$ & $H_{d}$ & $H^{'}_{d}$ \\
\hline
$Sp(1)$ &  $\ytableausetup{textmode, centertableaux, boxsize=0.6em}
\begin{ytableau}
 \\
\end{ytableau}$  &  -  & - & - & - &- \\
\hline
$Z_{4R}$ &  +1  &  0  & $x$ & $x$ & $-x$ &$-x$ \\
\hline
$Z_{4}$ &  +1  &  +2  & \multicolumn{2}{c|}{$H_u\leftrightarrow H_u' $}  & \multicolumn{2}{c|}{$H_d\leftrightarrow H_d' $} \\
\hline
\end{tabular}
\caption{Charge assignment for the chiral superfields.} 
\label{table1} 
\end{table}

We consider $Sp(1)$ gauge theory with $N_{F}=4$ matter fields $Q_{i}$ with $i$ being the flavor index running from 1 to 4. We also introduce singlet fields $S_{ij}$ with anti-symmetric indices. As we will see, a linear combination of $S_{ij}$ is a SUSY-breaking field. We may impose an anomaly-free discrete $R$-symmetry $Z_{4R}$ and  discrete flavor symmetry $Z_{4}$.  For the vanishing mixed anomaly of $Z_{4R}-[SU(2)_{L}]^{2}$ and $Z_{4R}-[SU(3)_{c}]^{2}$ within the MSSM, $R[H_{u}H_{d}]=0$ modulo 4 is required~\cite{Evans:2011mf}. The same applies for the mirror sector. Therefore, we choose $R[H_{u}H_{d}]=R[H^{'}_{u}H^{'}_{d}]=0$. In Table.~\ref{table1}, we show the charge assignment for the SUSY-breaking sector and ($H_{u}$,$H_{d}$) and ($H^{'}_{u}$,$H^{'}_{d}$). 
Under $Z_4$, $S_{ij}$ is transformed into $-S_{ij}$ and the MSSM sector is transformed into the mirror sector, so the $Z_4$ symmetry may be identified with the exchange symmetry in the TH model.

The superpotential of the SUSY-breaking sector above the dynamical scale $\Lambda_*$ is
\be
W\supset\lambda_{ijk\ell}S_{ij}Q_{k}Q_{\ell}\,,
\label{eq:Weff}
\ee
where
$\lambda_{ijk\ell}$ is a dimensionless coupling constant. For simplicity, we take $\lambda_{ijk \ell}=\lambda \delta_{ik} \delta_{j\ell}$, which preserves global $SU(4)$ symmetry. Below the dynamical scale, there arises a deformed moduli constraint ${\rm Pf}[Q_{i}Q_{j}]=\Lambda_{*}^{4}$~\cite{Seiberg:1994bz}. The SUSY-breaking sector can be described by the effective theory of the meson fields and the quantum moduli constraint can be equivalently rewritten as ${\rm Pf}[M_{ij}]=\Lambda_{*}^{2}/(4\pi)^{2}$, where six meson fields are defined via $M_{ij}\equiv Q_{i}Q_{j}/(4\pi\Lambda_{*})$. The mesons are sixplet of the global $SU(4) \simeq SO(6)$, and the global rotation allows us to take a basis where $\langle M_{1}\rangle=\Lambda_{*}/(4\pi)$ and $\langle M_{a}\rangle=0$ ($a=2-6$).
We denote the corresponding basis of $S_{ij}$ fields as $S_a$. We denote $M_{1}$ and $S_1$ simply as $M$ and $S$ from this point forward. 

After integrating out $M$ with the constraint $MM+M_{a}M_{a}=\Lambda_{*}^{2}/(16\pi^{2})$,
the SUSY-breaking sector can be effectively described by~\cite{Chacko:1998si} 
\begin{align}
K_{\rm eff}=&S^{\dagger}S+S_{a}^{\dagger}S_{a}+M_{a}^{\dagger}M_{a}+...\,, \nonumber \\
W_{\rm eff}=&\frac{\lambda}{4\pi}(\Lambda_{*}S\sqrt{\Lambda_{*}^{2}/(16\pi^{2})-M_{a}M_{a}}+\Lambda_{*}S_{a}M_{a})\,.
\label{eq:eff}
\end{align}
Here the factors of $4\pi$s in $W_{\rm eff}$ are determined in accordance with the naive dimensional analysis~\cite{Luty:1997fk,Cohen:1997rt}.

The $F$ terms of $S$ and $S_a$ are
\ba
\frac{\partial W_{\rm eff}}{\partial S}&=&\frac{\lambda}{4\pi}\Lambda_{*}\sqrt{\Lambda_{*}^{2}/(16\pi^{2})-M_{a}M_{a}}\,,\cr\cr\frac{\partial W_{\rm eff}}{\partial S_{a}}&=&\frac{\lambda}{4\pi}\Lambda_{*}M_{a}\,.
\label{eq:Fterms}
\ea
We see that $\partial W_{\rm eff}/\partial S=0$ and $\partial W_{\rm eff}/\partial S_{a}=0$ cannot be satisfied simultaneously, which implies SUSY-breaking. At the minimum of the potential, ${\rm Im}M_a=0$, while ${\rm Re}M_a$ is not fixed. They correspond to the Nambu-Goldstone bosons associated with $SO(6)/SO(5)$. They can be lifted by breaking $SO(6)$ symmetry with different Yukawa couplings in Eq.~(\ref{eq:Weff}) without disturbing the SUSY-breaking dynamics~\cite{Chacko:1998si}. In particular, by taking the Yukawa coupling of $S$ to be smaller than that of $S_a$, it is ensured that $\langle M\rangle=\Lambda_{*}/(4\pi)$ and $\langle M_{a}\rangle=0$.
From the $F$-term of $S$, $F_{S}=\lambda\Lambda_{*}^2/(4\pi)^2$, and the vanishing vacuum constant, the gravitino mass is given by
\be
m_{3/2}=\frac{\lambda}{\sqrt{3}M_{P}}\left(\frac{\Lambda_{*}}{4\pi}\right)^{2}\,.
\label{eq:m32}
\ee

The $F$-term conditions of $M_a$ give
\be
\frac{\partial W_{\rm eff}}{\partial M_{a}}\simeq\frac{\lambda}{4\pi}\frac{\Lambda_{*} SM_{a}}{\sqrt{(\Lambda_{*}^{2}/16\pi^{2})-M_{a}M_{a}}}-\frac{\lambda}{4\pi}\Lambda_{*}S_{a}=0\quad\Rightarrow\quad S_{a}\simeq\frac{SM_{a}}{\sqrt{(\Lambda_{*}^{2}/16\pi^{2})-M_{a}M_{a}}}\,.
\label{eq:Sa}
\ee
So each $S_a$ is fixed for a given $M_a$.
For our choice of the basis with $\langle M\rangle=\Lambda_{*}/(4\pi)$ and $\langle M_{a}\rangle=0$ ($a=2-6$), we have $\langle S_{a}\rangle=0$.

$S$ is massless at the tree-level~\cite{Arkani-Hamed:1997lye}.
Once we integrate out heavy fields at the energy scale below $\Lambda_{*}$, we obtain  
\be
W_{\rm eff}\supset\lambda\left(\frac{\Lambda_{*}}{4\pi}\right)^{2}S\,.
\label{eq:Weff2}
\ee
This form of $W_{\rm eff}$ applies to all range of $S$; for $\lambda S>\Lambda_{*}$, the theory becomes a pure gauge theory with a $S$-dependent gauge coupling and gaugino condensation induces the effective superpotential in Eq.~(\ref{eq:Weff2}).

Let us discuss the potential of $S$ given by one-loop corrections. For $\lambda S<\!\!<\Lambda_{*}$, the  K\"{a}hler potential of $S$ is given by~\cite{Chacko:1998si,Ibe:2006am}
\be
K_{\rm eff}\supset |S|^{2}-\frac{\eta\lambda^{2}}{4}\frac{|S|^{4}}{\Lambda_{*}^{2}}+ \cdots,
\label{eq:KforS}
\ee
where $\eta$ is a constant that is expected to be ${\cal O}(1)$. 
From Eqs.~(\ref{eq:Weff2}) and (\ref{eq:KforS}), the potential of $S$ is
\be
V(S)\simeq\left(\frac{\lambda\Lambda_{*}^{2}}{(4\pi)^{2}}\right)^{2}\left(1+\eta\lambda^{2}\frac{|S|^{2}}{\Lambda_{*}^{2}}\right)\quad({\rm for}\,\,\,\lambda S\,{\rm<\!\!<\Lambda_{*}})\,.
\label{eq:Spotential1}
\ee
Near the origin of $S$ field space, the mass of $S$ is
\be
m_{S}^{2}\simeq\frac{\eta\lambda^{3}}{(4\pi)^{2}}\sqrt{3}M_{P}m_{3/2}\,.
\label{eq:mS}
\ee
Note that there is no linear term of $S$ in the K\"{a}hler potential and no tilt of the potential that could destabilize $S$ from around the origin. This is in contrast to the MSSM case, where a linear term arises at quantum level even if $Z_2$ symmetry is imposed to the tree-level potential, and the Polonyi problem is reintroduced~\cite{Ibe:2006am}.

The condensation of $QQ$ breaks $Z_4 \times Z_4$ down to $Z_{4R}$, under which $S$ has a charge of $2$. $Z_{4R}$ is further broken down to $Z_{2R}$ by the non-zero gravitino mass. There is no residual symmetry under which $S$ is charged and we expect a non-zero vev of $S$.
Indeed, the effective superpotential in Eq.~(\ref{eq:Weff2}) and the supergravity effect induce a tadpole term of $S$, $V(S)\supset\lambda m_{3/2}\Lambda_{*}^2 S / (4\pi)^2$. The balance between the tadpole term and the mass term in Eq.~(\ref{eq:Spotential1}) gives
\be
\langle S\rangle\simeq\frac{(4\pi)^{2}}{2\lambda^{3}\eta}m_{3/2}\,,
\label{eq:Svev}
\ee
which is much smaller than $M_P$.
Also, as long as $\lambda > \Lambda_* / M_P$, $\lambda \langle S\rangle$ is indeed smaller than $\Lambda_*$ and the above computation is consistent.

The non-zero vev of $S$ may result in a non-zero $F$ term of the meson field $M$. To see that this is possible, we introduce a Lagrange multiplier $X$ for the deformed moduli constraint, which may be understood as a glueball field,
\be
W \supset 4\pi X\left(M^2 - \frac{\Lambda_*^2}{16\pi^2}\right).
\ee
The $F$ term of $M$ is
\be
-\langle F_{M}^*\rangle=\frac{\lambda}{4\pi}\Lambda_{*}\langle S\rangle+8\pi\langle X\rangle\langle M\rangle=\frac{(4\pi)}{2\lambda^{2}\eta}\Lambda_{*}m_{3/2}+2\Lambda_{*}\langle X\rangle\,.
\label{eq:FM}
\ee
It is in principle possible that the two terms cancel with each other, but we expect that the $F$ term is of the same order as the first term. This has an implication to the possibility where the $\mu$ term of the Higgs comes from the meson condensation as discussed in Appendix~\ref{sec:appendixB}.

%%%%%%%%%%%%%%%%%%%%%%%%%%%%%%%%%%%%%%%%%%%%
\subsection{Cosmological evolution of the Polonyi field}
\label{sec:cosmology}
%%%%%%%%%%%%%%%%%%%%%%%%%%%%%%%%%%%%%%%%%%%%

The gravitino mass in the range $m_{3/2}>\mathcal{O}(0.1-1){\rm TeV}$ gives the lower bound on the SUSY-breaking scale via Eq.~(\ref{eq:m32}),
\be
m_{3/2}>\mathcal{O}(0.1-1){\rm TeV}\quad\rightarrow\quad\Lambda_{*}\gtrsim\mathcal{O}(10^{10}){\rm GeV}\,.
\label{eq:Lambdastar}
\ee
We note that there can be two sources of domain wall problem: the spontaneous breaking of $Z_{4}$ and $Z_{4R}$ when the $Sp(1)$ gauge theory becomes strongly coupled and the $\langle QQ\rangle$ condensation forms. To avoid this problem, we require 
\be
\Lambda_{*}\gtrsim H_{\rm inf}\,,
\label{eq:DW}
\ee
where $H_{\rm inf}$ is the Hubble expansion rate during the inflation.\footnote{One may refer to \cite{Choi:2022fce} for a potential gravitational wave signature of the model with the discrete $R$-symmetry and the strong dynamics for the dynamical SUSY-breaking.}

The mass of the Polonyi field is given by Eq.~(\ref{eq:mS}).
Let us first discuss the case with $H_{\rm inf}>m_{S}$. During the inflation, $S$ field is fixed at the minimum of Hubble-induced potential. We assume that the Hubble induced mass is positive so that the origin, $S=0$, is the minimum.
Note that this is possible because of the $Z_{2}$-odd nature of $S$.

After inflation ends, the Hubble-induced mass decreases and when $H$ becomes as small as $m_{S}$ in Eq.~(\ref{eq:mS}), $S$ field starts the coherent oscillation around the minimum of $V(S)$ given by Eq.~(\ref{eq:Svev}). The amplitude of the oscillation is comparable to $\langle S\rangle=\mathcal{O}((4\pi)^2m_{3/2})$, which is much smaller than $M_{P}$. Therefore, the initial energy density of $S$ field ($\rho_{S,{\rm ini}}$) gets reduced by a factor $(m_{3/2}^{2}M_{P}^{2})/(m_{S}^{2}\langle  S\rangle^{2})\simeq(M_{P}/m_{3/2})\times(4\pi)^{-2}$ in our model in comparison with the case with neutral $S$. 

The condensate of the Polonyi field will mainly decay to a pair of gravitinos, and the BBN constrains the number density $n_{3/2}$ of gravitinos normalized by the entropy density $s$, $Y_{3/2}\equiv n_{3/2}/s$.  For $m_{3/2}=\mathcal{O}(0.1-1){\rm TeV}$, $m_{S}=\mathcal{O}(10^{9}){\rm GeV}$, and the decay rate of $S$ is $\Gamma_{S}\simeq m_{S}^{5}/(300m_{3/2}^{2}M_{P}^{2})=\mathcal{O}(100){\rm GeV}$~\cite{Ibe:2006am}.
Unless the reheating temperature of the universe is above $\sqrt{\mathcal{O}(100){\rm GeV}M_{P}}\simeq10^{10}{\rm GeV}$, the Polonyi field decays before the completion of reheating.
Note that $n_S/\rho_{\rm inf}$ is conserved after the beginning of the Polonyi oscillation but before the completion of reheating, where $\rho_{\rm inf}$ is the energy density of the inflaton. Using this and $m_S\sim H$ at the beginning of the oscillation, we obtain
\be
Y_{3/2}=\frac{n_{3/2}}{s}\simeq\frac{2n_{S}}{\frac{4\rho}{3T_{\rm rh}}}=\frac{T_{\rm rh}\langle S\rangle^{2}}{2m_{S}M_{P}^{2}}\sim (4\pi)^{5}\frac{T_{\rm rh}m_{3/2}^{3/2}}{M_{P}^{5/2}} \sim10^{-27}\times\left(\frac{T_{\rm rh}}{10^{9}{\rm GeV}}\right)\left(\frac{m_{3/2}}{1{\rm TeV}}\right)^{3/2}\,.
\label{eq:Y32}
\ee
This is well-below the BBN constraint $Y_{3/2} <10^{-16}$~\cite{Kawasaki:2004qu} and thus the Polonyi problem is completely resolved in our model.

For $H_{\rm inf}<m_{S}$, 
the oscillation amplitude of $S$  is even smaller and the Polonyi problem is also absent.

%%%%%%%%%%%%%%%%%%%%%%%%%%%%%%%%%%%%%%%%%%%%
\section{Summary and discussion}
\label{sec:conclusion}
%%%%%%%%%%%%%%%%%%%%%%%%%%%%%%%%%%%%%%%%%%%%
The SUSY TH theories are appealing frameworks with double protection of the electroweak scale by SUSY and an approximate global symmetry. However, as in the MSSM,
if SUSY TH models are embedded into the simplest mediation scheme of SUSY breaking, namely, gravity mediation, they encounter the possibility of too large energy density of the Polonyi field $S$.
In this work, we considered the case in which $S$ transforms as $S\rightarrow-S$ under the discrete $Z_{2}$ symmetry exchanging the SM and mirror sectors. The Polonyi field can couple to gauge multiplets to give tree-level gaugino masses of the order of scalar masses.   Because of the $Z_{2}$-odd nature, the symmetry enhanced point of the $S$ field space, namely, the origin, is clearly identified, which results in great reduction of the amplitude of the oscillations of the Polonyi field in the early universe and the Polonyi problem is resolved in our setup.

The SUSY breaking by $S$ spontaneously breaks the $Z_2$ symmetry and induces $Z_2$-breaking soft masses and the $\mu$ term.
This can explain the required $Z_2$-breaking in the Higgs potential.
Also, the degeneracy of the masses of supersymmetric particles with their twin partners is resolved.  
As discussed in Sec.~\ref{sec:model}, the model predicts the splitting of masses of the two sector by quantum corrections of $\mathcal{O}(0.1-1)\%$, ${\cal O}(10)\%$, $\mathcal{O}(1)\%$ times $(M_{\rm cut}/M_*)$ for gauginos, squarks, and sleptons, respectively. 
We leave the study of the phenomenological implication of the mass splitting for future work. 

In this paper, we focused on a $Z_2$-odd Polonyi field in supersymmetric Twin Higgs models motivated by the little hierarchy problem. A similar idea can also work for another scenarios with $Z_2$ symmetry that exchanges gauge fields with their partners. For example, mirror $Z_2$ symmetry is also motivated from the heavy QCD axion scenario that relaxes the quality problem~\cite{Rubakov:1997vp,Berezhiani:2000gh,Fukuda:2015ana,Hook:2019qoh}. The Polonyi field may be also $Z_2$-odd in such a scenario. Another possible scenario is a solution to the strong CP problem based on a UV $SU(3)_c\times SU(3)_c' \times SU(2)_L \times SU(2))L'\times U(1)(\times U(1)')$ gauge symmetry and a parity symmetry~\cite{Barr:1991qx}. In this case, however, $S$ should transform as $S\rightarrow S^\dag$ under the parity symmetry and the real part of $S$ is not $Z_2$-odd. We need to add an extra $Z_2$ symmetry under which the real part of $S$ is odd. We leave the survey of various possibilities of a  $Z_2$-odd Polonyi field to future work.

%%%%%%%%%%%%%%%%%%%%%%%%%%%%%%%%%%%%%%%%%%%%
\section*{Acknowledgments}
G.C. would like to acknowledge the Mainz Institute for Theoretical Physics (MITP) of the Cluster of Excellence PRISMA+(Project ID 39083149), for its hospitality and its partial support during the completion of this work.
%%%%%%%%%%%%%%%%%%%%%%%%%%%%%%%%%%%%%%%%%%%%

\appendix
%%%%%%%%%%%%%%%%%%%%%%%%%%%%%%%
\section{Useful equations for computing soft masses} 
\label{sec:appendixA}
The following formulae are taken from~\cite{Martin:1997ns}.
\subsection{RGEs for sfermion and Higgs masses}
\label{sec:appendixA1}
\setcounter{equation}{0}
%%%%%%%%%%%%%%%%%%%%%%%%%%%%%%%

\begin{align}
16\pi^{2}\frac{d}{dt}m_{\tilde{Q}_{L}}^{2}&\supset-\frac{32}{3}g_{3}^{2}|M_{3}|^{2}+2|A^{q}|^{2}\,,\nonumber \\
16\pi^{2}\frac{d}{dt}m_{\tilde{u}_{R}}^{2}&\supset-\frac{32}{3}g_{3}^{2}|M_{3}|^{2}+4|A^{q}|^{2}\,,\nonumber \\
16\pi^{2}\frac{d}{dt}m_{\tilde{d}_{R}}^{2}&\supset-\frac{32}{3}g_{3}^{2}|M_{3}|^{2}+4|A^{q}|^{2}\,,\nonumber \\
16\pi^{2}\frac{d}{dt}m_{\tilde{L}_{L}}^{2}&\supset-6g_{2}^{2}|M_{2}|^{2}+2|A^{\ell}|^{2}\,,\nonumber \\
16\pi^{2}\frac{d}{dt}m_{\tilde{e}_{R}}^{2}&\supset-\frac{24}{5}g_{1}^{2}|M_{1}|^{2}+4|A^{\ell}|^{2}\,,\nonumber \\
16\pi^{2}\frac{d}{dt}m_{H_{u}}^{2}&\supset-6g_{2}^{2}|M_{2}|^{2}+6|A^{t}|^{2}+6|y_{t}|^{2}(m_{H_{u}}^{2}+m_{t}^{2}+m_{t^{c}}^{2})\,,\nonumber \\
16\pi^{2}\frac{d}{dt}m_{H_{d}}^{2}&\supset-6g_{2}^{2}|M_{2}|^{2}+6|A^{b}|^{2}+6|y_{b}|^{2}(m_{H_{d}}^{2}+m_{b}^{2}+m_{b^{c}}^{2})\,,
\label{eq:RGEsfermion}
\end{align}
where $t=d\log\mu$ and each subscript for $m_{\tilde{f}}^{2}$ denotes quark and lepton doublets and singlets.

\subsection{RGEs for $A$, $\mu$ and $b$ parameters}
\label{sec:appendixA2}

\ba
16\pi^{2}\frac{d}{dt}A^{t}&\supset&A^{t}(18y_{t}^{*}y_{t}-\frac{16}{3}g_{3}^{2}-3g_{2}^{2})+y_{t}\frac{32}{3}g_{3}^{2}M_{3}
\,,\cr\cr
16\pi^{2}\frac{d}{dt}\mu&\supset&\mu\left[3y_{t}^{*}y_{t}-3g_{2}^{2}-\frac{3}{5}g_{1}^{2}\right]\,,\cr\cr
16\pi^{2}\frac{d}{dt}b&\supset&b(3y_{t}^{*}y_{t}-3g_{2}^{2})+6\mu g_{2}^{2}M_{2}\,,
\label{eq:RGEsAb}
\ea

\subsection{Anomalous dimensions}
\label{sec:appendixA3}
\ba
\gamma_{H_{u}}&=&\frac{1}{16\pi^{2}}\left[3y_{t}^{*}y_{t}-\frac{3}{2}g_{2}^{2}-\frac{3}{10}g_{1}^{2}\right]\,,\cr\cr
\gamma_{Q_{3}}&=&\frac{1}{16\pi^{2}}\left[y_{t}^{*}y_{t}+y_{b}^{*}y_{b}-\frac{8}{3}g_{3}^{2}-\frac{3}{2}g_{2}^{2}-\frac{1}{30}g_{1}^{2}\right]\,,\cr\cr
\gamma_{u_{3}^{c}}&=&\frac{1}{16\pi^{2}}\left[2y_{t}^{*}y_{t}-\frac{8}{3}g_{3}^{2}-\frac{8}{15}g_{1}^{2}\right]\,.
\label{eq:anom}
\ea

%%%%%%%%%%%%%%%%%%%%%%%%%%%%%%%
\section{$\mu$ and $b$ terms from the hidden strong dynamics} 
\label{sec:appendixB}
\setcounter{equation}{0}
%%%%%%%%%%%%%%%%%%%%%%%%%%%%%%%
The charge assignment specified in Table.~\ref{table1} allows the following  superpotential terms,
\be
W\supset\frac{h_{ij}}{M_{P}}Q_{i}Q_{j}\left( H_{u}H_{d} -H'_{u}H'_{d}\right),
\label{eq:WHuHd}
\ee
where $h_{ij}$ is a dimensionless coupling constant. In the confined phase of $Sp(1)$, thanks to the deformed moduli constraint ${\rm Pf}[Q_{i}Q_{j}]=\Lambda_{*}^{4}$, Eq.~(\ref{eq:WHuHd}) leads to 
\be
W_{\rm eff}\supset\frac{h}{M_{P}}\left(\frac{\Lambda_{*}}{4\pi}\right)^{2} \left( H_{u}H_{d} - H'_{u}H'_{d} \right),
\label{eq:WeffHuHd}
\ee
where we suppressed the flavor indices of the coupling constant for simplicity. From Eqs.~(\ref{eq:WeffHuHd}) and (\ref{eq:m32}), one can read the following contribution to $\mu$ term from the coupling of $Sp(1)$ quark fields to the operator $H_{u}H_{d}$
\be
\mu=\frac{\sqrt{3}h}{\lambda}m_{3/2}\,.
\label{eq:muWeff}
\ee

In terms of the meson field $M$, Eq.~\eqref{eq:WeffHuHd} reads
\be
W_{\rm eff}\supset\frac{h\Lambda_{*}}{4\pi M_{P}}M \left( H_{u}H_{d}- H'_{u}H'_{d} \right)\,,
\label{eq:MHuHd}
\ee
The F term of $M$ is expected to be non-zero. Using Eq.~\eqref{eq:FM}, we find
\be
b = - b' \sim h \frac{(4\pi)^2}{\lambda^3} m_{3/2}^2.
\ee
For perturbative $\lambda$, if $h = {\cal O}(1)$, $|b| \gg m_{3/2}^2$ and a stable electroweak vacuum does not exist, so we need $h \ll 1$. This provides another way to suppress the $Z_2$ breaking in the Higgs potential while $\tan \beta = 2-3$. The $Z_2$-odd $b$ terms from the meson field may be comparable to $m_{H_u}^2$ and $m_{H_d}^2$ to ensure $\tan \beta = 2-3$. The $\mu$ term generated from the meson field is too small, but we may obtain $\mu$ from $R$ symmetry breaking, as in Eq.~\eqref{eq:btree}. It is generically $Z_2$-breaking, but as long as $\mu^2 \ll m_{H_u}^2,m_{H_d}^2$, the $Z_2$-breaking is suppressed.

%%%%%%%%%%%%%%%%%%%%%%%%%%%%%%%%%%%%%%%%%%%%%%%%%%
\bibliography{main}
\bibliographystyle{jhep}
%%%%%%%%%%%%%%%%%%%%%%%%%%%%%%%%%%%%%%%%%%%%%%%%%%

\end{document}